\documentclass[11pt]{article}

\usepackage[preprint]{acl}

\usepackage{times}
\usepackage{latexsym}

\usepackage[T1]{fontenc}

\usepackage[utf8]{inputenc}

\usepackage{microtype}

\usepackage{inconsolata}

\usepackage{graphicx}

\usepackage{booktabs}
\usepackage{colortbl}
\usepackage{pifont}
\usepackage{multirow}
\usepackage{tabularx}
\usepackage{amsmath}
\usepackage{amssymb}
\usepackage{mathtools}
\usepackage[most]{tcolorbox}
\usepackage[capitalise,noabbrev,nameinlink]{cleveref}

\newcommand{\GoodMark}{\textcolor{green!60!black}{\ding{51}}}
\newcommand{\BadMark}{\textcolor{red!70!black}{\ding{55}}}

\definecolor{lightgray}{gray}{0.9}
\definecolor{lggray}{gray}{0.95}
\definecolor{promptAccent}{RGB}{15,56,112}
\definecolor{promptBack}{RGB}{245,248,255}
\definecolor{text2svgAccent}{RGB}{124,58,237}
\definecolor{text2svgBack}{RGB}{245,240,255}
\definecolor{sketch2svgAccent}{RGB}{16,122,28}
\definecolor{sketch2svgBack}{RGB}{239,250,241}
\definecolor{svgeditAccent}{RGB}{234,88,12}
\definecolor{svgeditBack}{RGB}{255,246,235}
\definecolor{svgcaptionAccent}{RGB}{71,85,105}
\definecolor{svgcaptionBack}{RGB}{248,250,252}

\newtcolorbox[auto counter]{PromptBox}[2][]{%
  enhanced,
  breakable,
  colback=promptBack,
  colframe=promptAccent,
  coltitle=white,
  fonttitle=\bfseries\small\ttfamily,
  fontupper=\ttfamily\footnotesize,
  title={\ifx&#2&Prompt~\thetcbcounter\else#2\fi},
  boxed title style={
    boxrule=0pt,
    arc=2pt,
    outer arc=2pt,
    top=2pt,
    bottom=2pt,
    left=6pt,
    right=6pt,
  },
  boxrule=0.75pt,
  arc=2pt,
  boxsep=5pt,
  before skip=6pt,
  after skip=6pt,
  borderline west={2pt}{0pt}{promptAccent},
  drop shadow southeast,
  sharp corners=south,
  before upper={\vspace{2pt}},
  #1
}

\title{VectorGym: A Multi-Task Benchmark for \\
SVG Code Generation, Sketching and Editing}

\author{
  \textbf{Joan Rodriguez\textsuperscript{1,2,14}\thanks{Equal contribution.}},
  \textbf{Haotian Zhang\textsuperscript{3,14}\footnotemark[1]},
  \textbf{Abhay Puri\textsuperscript{4}},
  \textbf{Haoran Dai\textsuperscript{5,14}},
  \\
  \textbf{Tianyang Zhang\textsuperscript{6,14}},
  \textbf{Meng Lin\textsuperscript{8,14}},
  \textbf{Rishav Pramanik\textsuperscript{7}},
  \textbf{Xiaoqing Xie\textsuperscript{9,14}},
  \textbf{Marco Terral Rodriguez\textsuperscript{10,14}},
  \\
  \textbf{Darsh Kaushik\textsuperscript{1,14}},
  \textbf{Aly Shariff\textsuperscript{11}},
  \textbf{Perouz Taslakian\textsuperscript{4}},
  \textbf{Spandana Gella\textsuperscript{4}},
  \\
  \textbf{Sai Rajeswar\textsuperscript{4,1}},
  \textbf{David Vazquez\textsuperscript{4}},
  \textbf{Christopher Pal\textsuperscript{4,1,12,13}},
  \textbf{Marco Pedersoli\textsuperscript{1,2}}
  \\
  \textsuperscript{1}Mila, Quebec AI Institute,
  \textsuperscript{2}\'{E}TS Montr\'eal,
  \textsuperscript{3}Columbia University,
  \textsuperscript{4}ServiceNow Research,
  \\
  \textsuperscript{5}Illinois Institute of Technology,
  \textsuperscript{6}University of Massachusetts Amherst,
  \textsuperscript{7}Stony Brook University,
  \\
  \textsuperscript{8}University of Illinois Urbana-Champaign,
  \textsuperscript{9}Minzu University of China,
  \textsuperscript{10}Computer Vision Center,
  \\
  \textsuperscript{11}University of Waterloo,
  \textsuperscript{12}Polytechnique Montr\'eal,
  \textsuperscript{13}Canada CIFAR AI Chair,
  \textsuperscript{14}QuiverAI
  \\
  \small{
    \textbf{Correspondence:} \href{mailto:juan.rodriguez@mila.quebec}{juan.rodriguez@mila.quebec}
  }
}

\newcommand{\vectorgymteasercaption}{VectorGym is a suite of human-authored datasets covering Sketch2SVG (\textbf{VG-Sketch}), SVG Editing (\textbf{VG-Edit}), Text2SVG (\textbf{VG-Text}), and SVG Captioning (\textbf{VG-Cap}). Unlike prior benchmarks, it is built from diverse real-world SVGs sourced from GitHub. Human experts annotate each SVG by hand-drawing sketches, creating complex edits, and writing detailed text descriptions, which are further cleaned and adapted into instruction-style prompts at varying levels of detail. We evaluate state-of-the-art models in VectorGym.}

\newcommand{\vectorgymteaser}{%
  \vskip 0.2in
  {\centering
    \includegraphics[width=0.94\textwidth]{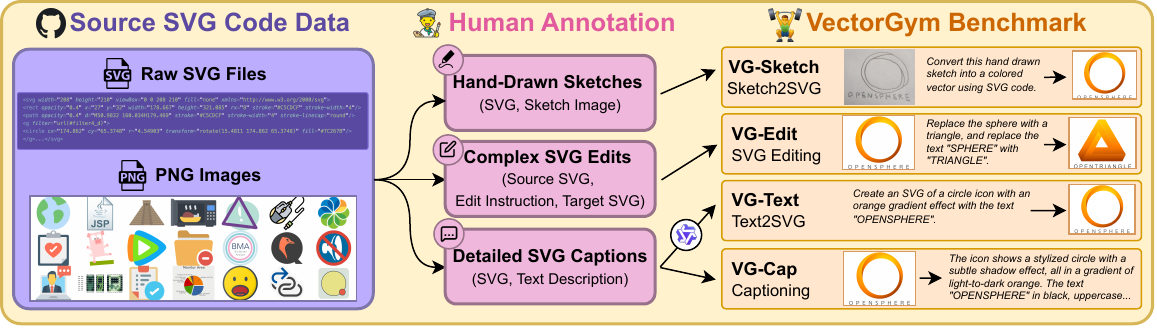}\par
  }
  \vskip 0.025in
  \refstepcounter{figure}\label{fig:teaser}%
  {\footnotesize\textbf{Figure~\thefigure: Overview of VectorGym.} \vectorgymteasercaption\par}
}

\makeatletter
\AtBeginDocument{%
\def\@maketitle{\vbox{\hsize\textwidth
 \linewidth\hsize \vskip 0.08in \centering
 {\Large\bfseries \@title \par} \vskip 0.12in
 {\def\and{\unskip\enspace{\rmfamily and}\enspace}%
  \def\And{\end{tabular}\hss \egroup \hskip 1in plus 2fil
           \hbox to 0pt\bgroup\hss \begin{tabular}[t]{c}\bfseries}%
  \def\AND{\end{tabular}\hss\egroup \hfil\hfil\egroup
          \vskip 0.25in plus 1fil minus 0.125in
           \hbox to \linewidth\bgroup\large \hfil\hfil
             \hbox to 0pt\bgroup\hss \begin{tabular}[t]{c}\bfseries}%
  \hbox to \linewidth\bgroup\large \hfil\hfil
    \hbox to 0pt\bgroup\hss
  \outauthor
   \hss\egroup
    \hfil\hfil\egroup}
  \vectorgymteaser
  \vskip 0.2in
}}
}
\makeatother

\begin{document}
\maketitle

\begin{abstract}

We introduce VectorGym, a comprehensive benchmark suite for Scalable Vector Graphics (SVG) that spans generation from text and sketches, complex editing, and visual understanding. VectorGym addresses the lack of realistic, challenging benchmarks aligned with professional design workflows. Our benchmark comprises four tasks with expert human-authored annotations: the novel Sketch2SVG task (VG-Sketch); a new SVG editing dataset (VG-Edit) featuring complex, multi-step edits with higher-order primitives; Text2SVG generation (VG-Text); and SVG captioning (VG-Cap). Unlike prior benchmarks that rely on synthetic edits, VectorGym provides gold-standard human annotations that require semantic understanding and design intent. We also provide a multi-task reinforcement learning baseline that jointly optimizes across all four tasks using rendering-based rewards. This baseline, built on GRPO with curriculum learning, trains a Qwen3-VL 8B model that achieves state-of-the-art performance among open-source models, surpassing much larger models including Qwen3-VL 235B and matching GPT-4o. We also introduce a VLM-as-a-Judge metric for SVG generation, validated through human correlation studies. Our evaluation of frontier VLMs reveals significant performance gaps, positioning VectorGym as a rigorous framework for advancing visual code generation.

\end{abstract}

\section{Introduction}
\label{sec:intro}

Scalable Vector Graphics (SVG)~\citep{ferraiolo2000scalable, quint2003scalable} are widely used across the web, design tooling, and digital media. Unlike raster images~\citep{rodriguez2023figgen, rodriguez2023ocr, rombach2021highresolution}, SVGs are programs: their code exposes geometry, style, and structure, enabling precise editing, scalable rendering, and semantic manipulation. Evaluating models on SVG requires not only visual understanding but also reliable, syntax-aware code generation.

Despite rapid progress in Vision-Language Models (VLMs), existing evaluations of SVG generation remain limited in scope. Prior datasets often target icons or basic shapes, rely on synthetic programmatic edits, rarely assess sketch-conditioned generation nor provide human-authored gold labels~\citep{Rodriguez2025StarVector, wu2023iconshop, zhang2023pixelsexploringhumanreadablesvg, xing2025empowering, yang2025omnisvg, rodriguez2025rendering}. As a result, the field lacks a unified, realistic benchmark that simultaneously stresses visual understanding, vector generation and structured SVG code manipulation.

We introduce \textbf{VectorGym}, a new comprehensive multi-task benchmark for SVG generation and manipulation spanning four tasks: (1) \textbf{Sketch2SVG}(\textit{VG-Sketch}), converting rough sketches to clean vector code; (2) \textbf{SVG Editing}(\textit{VG-Edit}), applying natural-language edits to existing SVGs; (3) \textbf{Text2SVG} (\textit{VG-Text}), generating SVGs from text; and (4) \textbf{SVG Captioning} (\textit{VG-Cap}), describing SVG content. Our benchmark introduces the Sketch2SVG task and releases the first dataset of complex, human-authored SVG edits; all tasks use gold-standard human annotations.

\textbf{VectorGym} covers in-the-wild diversity, including icons, diagrams, emojis, fonts, logotypes, and complex illustrations, sourced from SVG-Stack~\citep{Rodriguez2025StarVector}. We pair this with careful human curation to ensure realistic and challenging task difficulty. We evaluate both proprietary and open-source frontier VLMs, providing a clear view of current capabilities and gaps. 

Our main contributions are:

\begin{enumerate}
\item We introduce a comprehensive, multi-task benchmark for real-world SVG code generation with gold-standard human annotations across all tasks;

\item We introduce the Sketch2SVG task and the first dataset of expert human-authored SVG edits with complex intent, involving rich primitives and non-trivial edits.

\item We provide a multi-task reinforcement learning baseline that jointly optimizes models across all four VectorGym tasks, achieving state-of-the-art performance among open-source models. In addition, we propose a task-specific VLM-as-a-judge evaluation suite for SVG generation, covering sketch, text, and editing tasks, and validate it through human correlation studies.

\item We provide extensive evaluation and analysis of current frontier VLMs across SVG generation tasks.
\end{enumerate}

\begin{table*}[t]
\centering
\caption{\textbf{SVG Datasets Comparison.} We compare similar datasets across annotation quality, task complexity, and data source. We report the SVG type, total samples, whether each dataset supports multiple tasks, the level of SVG primitive coverage beyond simple paths (i.e. including higher-order primitives, like circle, text, gradients or animation logic), and whether it includes annotations for edits and sketches, plus whether these annotations reach expert human quality with complex intent. \textbf{VectorGym} is built from real SVGs collected from GitHub (\textit{in-the-wild}), preserving original structure and primitive detail, and \textit{provides targets created by expert humans} with the goal of producing \textit{complex annotations} that capture semantic understanding, design intent, and multi-step reasoning.}

\label{tab:datasets-compact}
\setlength{\tabcolsep}{4pt}

\resizebox{\textwidth}{!}{
\begin{tabular}{@{}l c r c c c c c@{}}
\toprule
\textbf{Dataset} & \textbf{SVG Type} & \textbf{\# Samples} & \textbf{Multi-Task} & \textbf{Primitives} & \textbf{Edits} & \textbf{Sketches} & \textbf{Human} \\
\midrule
{SVGBench~\citep{Rodriguez2025StarVector}} 
  & {\textit{In-the-wild}} 
  & {10k} 
  & {\BadMark}
  & {\GoodMark} 
  & {\BadMark} 
  & {\BadMark} 
  & {\BadMark} \\
{VGBench~\citep{VGBench2024}} 
  & {Emojis} 
  & {9.5k} 
  & {\GoodMark}
  & {\BadMark} 
  & {\BadMark} 
  & {\BadMark} 
  & {\BadMark} \\
{SVGEditBench~\citep{nishina2024svgeditbench}} 
  & {Emojis} 
  & {1.6k} 
  & {\BadMark}
  & {\BadMark} 
  & {\GoodMark} 
  & {\BadMark} 
  & {\BadMark} \\
{SVGEditBenchV2~\citep{nishina2025svgeditbench}} 
  & {Emojis} 
  & {1.6k} 
  & {\BadMark}
  & {\BadMark} 
  & {\GoodMark} 
  & {\BadMark} 
  & {\BadMark} \\
{UniSVG~\citep{li2025unisvg}} 
  & {Icons} 
  & {525k} 
  & {\GoodMark}
  & {\BadMark} 
  & {\BadMark} 
  & {\BadMark} 
  & {\BadMark} \\
{SVGenius~\citep{chen2025svgenius}} 
  & {Icons} 
  & {100k} 
  & {\GoodMark}
  & {\BadMark} 
  & {\GoodMark} 
  & {\BadMark} 
  & {\BadMark} \\
\midrule
\textbf{{VectorGym (ours)}} 
  & {\textit{In-the-wild}} 
  & {7k} 
  & \cellcolor{gray!15}{\GoodMark}
  & \cellcolor{gray!15}{\GoodMark}
  & \cellcolor{gray!15}{\GoodMark}
  & \cellcolor{gray!15}{\GoodMark}
  & \cellcolor{gray!15}{\GoodMark} \\
\bottomrule
\end{tabular}
}
\end{table*}

\section{Related Work}
\label{sec:related}

\textbf{Vector Graphics Generation.} Classical vectorization methods based on shape-fitting algorithms~\citep{DiffVG2020,vtracer} struggle with complex tasks beyond image vectorization. Recent approaches introduce learning-based components, relying on latent variable models with differentiable rendering and attention architectures~\citep{Carlier2020DeepSVG,Cao2023SVGformer,Lopes2019SVGFonts}, as well as sketch abstraction~\citep{vinker2022clipasso} and text-conditioned SVG synthesis~\citep{jain2023vectorfusion}. However, these methods are still not general enough to support a wide range of SVG tasks.

\textbf{VLMs for SVG Generation.} Modern VLMs~\citep{openai2023gpt4, comanici2025gemini} can now produce structured code from visual inputs. StarVector~\citep{Rodriguez2025StarVector} frames SVG creation as a visual to code generation task, jointly testing visual understanding and program synthesis. Subsequent work further supports this direction~\citep{zhang2023pixelsexploringhumanreadablesvg, cai2023leveraging, yang2025omnisvg}.

\textbf{SVG Datasets and Benchmarks.} Foundational SVG datasets include DeepSVG icons~\citep{Carlier2020DeepSVG}, FIGR-8~\citep{Clouatre2019FIGR8}, and SVG-Stack~\citep{Rodriguez2025StarVector}. Several benchmarks address different SVG related tasks. UniSVG~\citep{li2025unisvg} unifies 525k SVGs for understanding and generation. VGBench~\citep{VGBench2024} aggregates multiple sources to evaluate image to SVG, text to SVG, and diagram code generation. SVGEditBench~\citep{nishina2024svgeditbench} and its V2 version~\citep{nishina2025svgeditbench} target instruction based editing using synthetic LLM generated edits or edits derived from similar SVGs. SVGenius~\cite{chen2025svgenius} covers a wide set of tasks, notably editing through algorithmic transform based operations.

Here we propose \textbf{VectorGym}, which focuses on edits created by humans following instructions that make the edits \textit{complex} and closer to the actions of real design professionals, requiring semantic understanding. We also introduce the novel Sketch2SVG task from human drawn sketches, and we collect human validated text captions that allow evaluation of both Text2SVG and SVG captioning on realistic, high difficulty edits. 
Sketch2SVG infers intent from noisy, underspecified sketches, whereas Text2SVG and conventional vectorization use explicit descriptions and complete renderings, respectively. VectorGym uniquely combines in-the-wild SVGs, sketch annotations, and expert human annotations, unlike prior editing benchmarks based on synthetic or programmatic transformations (Table~\ref{tab:datasets-compact}; Appendix~\ref{svg-datasets-appendix}).

\section[VectorGym Benchmark]{\raisebox{-0.1em}{\includegraphics[height=1em]{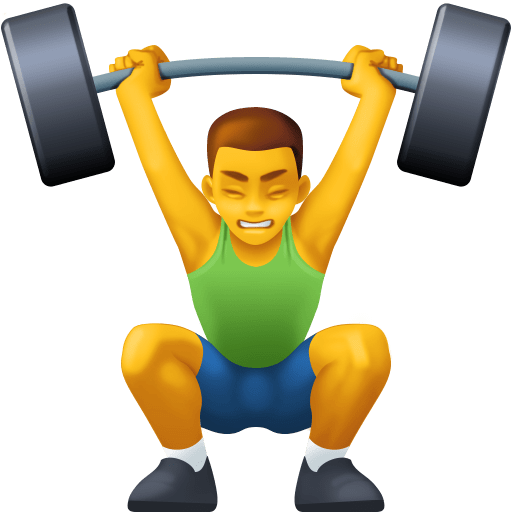}} VectorGym Benchmark}
\label{sec:vector-gym}

\begin{figure*}[t]
    \centering
    \includegraphics[width=1.0\linewidth]{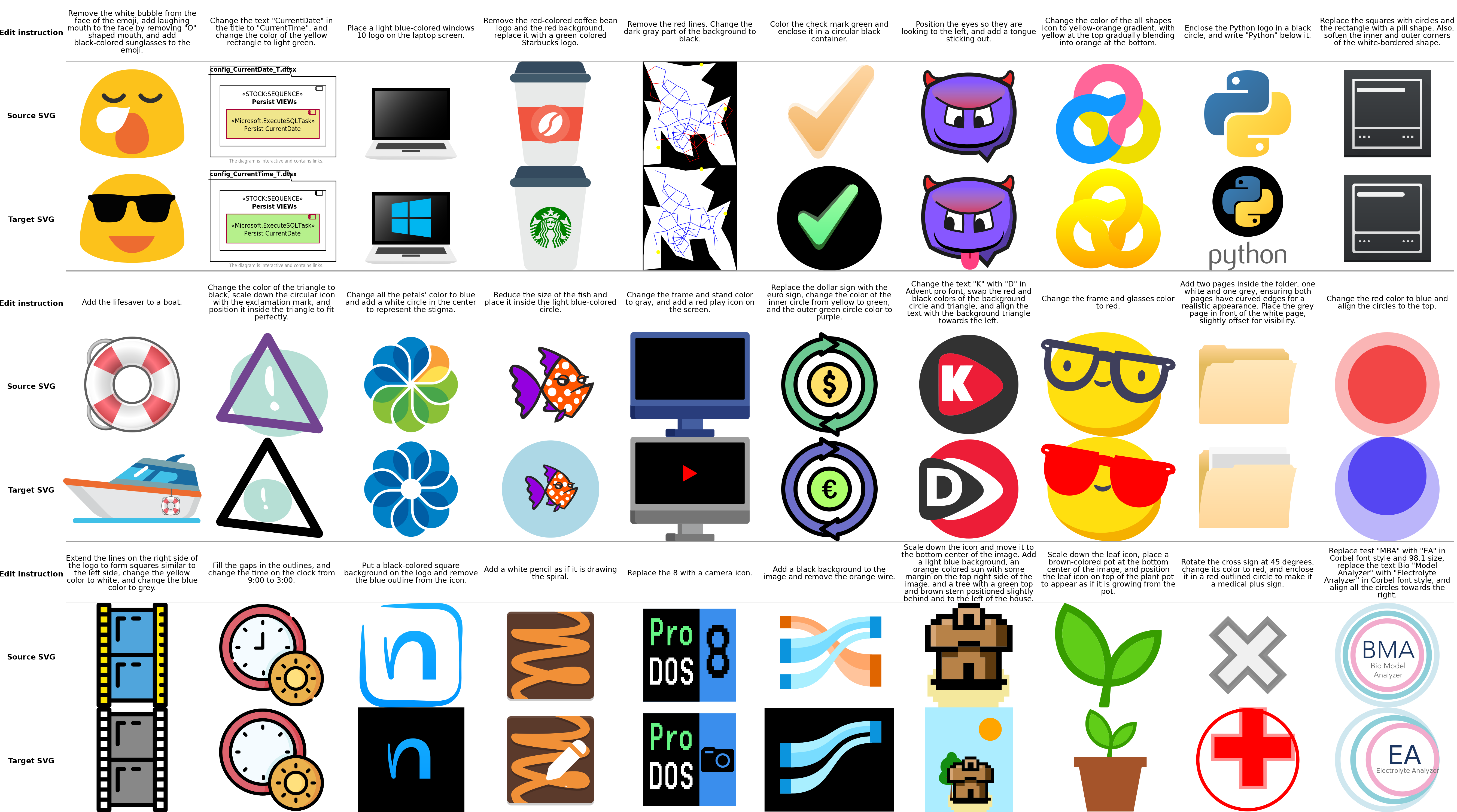}
    \caption{\textbf{Visualization of VectorGym Test Examples (Editing Task).} We randomly sample 21 examples from the test set, and show the editing instruction to perform, along with the source and target SVG. These examples are part of \textit{VG-Edit}.}
    \label{fig:vg-edit-samples}
\end{figure*}

VectorGym consists of four complementary tasks that comprehensively evaluate different aspects of SVG understanding and generation. Each task is designed to assess specific capabilities while contributing to a holistic understanding of visual2code generation performance.

SVG is well suited to this setting because its editable code exposes geometry, style, and object structure while mapping directly to rendered output. This enables joint assessment of visual fidelity, syntactic validity, and structured manipulation. Other visual-code formats, including Python plotting libraries, HTML/CSS, TikZ, and Canvas, provide higher-level or application-specific abstractions; dedicated benchmarks already cover document/code and TikZ generation~\citep{rodriguez2025bigdocs,belouadi2023automatikz}. We therefore focus on the general vector representation and leave systematic cross-format evaluation to future work.

\subsection{Task Definitions}

\textbf{Sketch2SVG Generation (\textit{VG-Sketch}).} This task evaluates the ability to convert rough, hand-drawn sketches into clean SVG code. Given a bitmap sketch image with approximate shapes and imperfect lines, models must generate SVG code that captures the essential geometric structure while producing a clean, scalable vector representation. This task tests spatial reasoning, shape recognition, and the ability to abstract from noisy visual input to structured geometric primitives.

\emph{Why Sketch2SVG?} Sketch2SVG captures an important early-stage design workflow: users often communicate visual intent through rough sketches rather than precise text or polished raster images. Unlike Text2SVG, where objects and attributes are explicitly specified, Sketch2SVG requires inferring semantic structure, topology, and spatial relations from noisy, incomplete, and ambiguous strokes. Unlike conventional image-to-SVG vectorization, the goal is not to trace a clean raster image but to abstract a rough human drawing into clean, valid, editable SVG code. Sketch2SVG therefore evaluates a distinct combination of visual abstraction, spatial reasoning, and structured code generation.

\textbf{SVG Editing (\textit{VG-Edit}).} In this task, models are given an SVG along with an editing instruction and must produce a new SVG with the specified edit applied. VG-Edit offers unprecedented \textit{complexity} in the challenge of SVG editing. Our editing instructions include deep understanding of the SVG syntax, requiring the use of complex primitives like texts, animations, or color gradients. It also requires multi-step reasoning and semantic understanding (See examples in Figure~\ref{fig:teaser} (right) and Figure~\ref{fig:vg-edit-samples}).

The challenge lies in correctly parsing the intent, identifying the relevant elements, and applying the transformation while preserving code validity, visual coherence, and the integrity of unmodified parts. Since instructions and targets were created by skilled human annotators, the edits are non-trivial, for example, adding new objects, modifying logo content or text, converting a pie chart to a bar chart, or changing facial expressions. This task evaluates both SVG structure understanding and the ability to follow complex editing instructions. Figure~\ref{fig:vg-edit-samples} shows examples from our test set. Unlike prior benchmarks~\cite{nishina2025svgeditbench, chen2025svgenius}, which focus on simple synthetic programmatic edits, \textit{VG-Edit introduces complex, high-difficulty editing scenarios annotated by human experts.}

\textbf{Text2SVG Generation (\textit{VG-Text}).} Given natural language descriptions of visual content, models must generate complete SVG code that accurately represents the described objects, scenes, or abstract concepts. Descriptions range from simple geometric shapes (``red circle with blue border'') to complex illustrations (``minimalist icon of a house with a tree''). This task tests creative generation capabilities and the ability to translate semantic concepts into precise geometric representations.

\textbf{SVG Captioning (\textit{VG-Cap}).} The inverse of Text2SVG generation, this task requires models to analyze existing SVG code and generate natural language descriptions that accurately capture the visual content, style, and key characteristics. High-quality captions should describe both the semantic content (``house icon'') and relevant visual properties (``minimalist style,'' ``blue and white color scheme''). This task evaluates SVG code comprehension and visual understanding.

\subsection{Dataset Construction}
Our datasets are built on a carefully curated SVG collection pipeline designed to ensure diversity across content types, complexity levels, and visual styles. We source high quality and diverse SVGs from the SVG Stack dataset~\citep{Rodriguez2025StarVector}, an established collection that includes icons, diagrams, emojis, fonts, logotypes, and complex illustrations. Since the original data was extracted from GitHub, it naturally reflects in-the-wild SVG code, including higher-order primitives such as text, gradients, polygons, and animations. This makes the dataset more representative of real design workflows and provides challenging examples for model development.

Our automatic curation builds on insights from prior SVG datasets~\citep{Carlier2020DeepSVG,Clouatre2019FIGR8,nishina2024svgeditbench,li2025unisvg,chen2025svgenius}. We extracted 7,000 candidate samples from the SVG Stack training split through multi stage filtering, including token length constraints (2k to 8k tokens to retain meaningful complexity), color entropy thresholding (normalized entropy greater than 0.55), and random subsampling followed by human visual inspection. After filtering, the final training set contains 6.5k samples. From these, we selected 100 samples to form our validation set, used for method tuning, in context learning, human evaluation, and metric design (see Appendix~\ref{app:vlmaj}). 
We applied the same pipeline to obtain 300 samples from the SVG-Stack test split and report all scores as point estimates.

\textbf{Human Annotation Process.} We partnered with two specialized data annotation vendors to produce high quality annotations across sketch and editing tasks. The process involved more than 20 annotators with diverse backgrounds and expertise in design, vector graphics, and coding. 
Annotators were provided with drawing tools, coding utilities, and curated SVG collections to perform edits and create sketches on different surfaces. They were specifically instructed to produce challenging edits, involving multi-step reasoning, and real design intent, and we iterated several times on these samples to validate their complexity and quality.
See Appendix~\ref{sec:annotation-details} for full details on the annotation methodology, quality assurance procedures, and complexity requirements.

\textbf{Complex Annotations.} In our setup, \textit{complex annotations} refer to human created editing instructions and corresponding SVG modifications that require things like deeper understanding of the SVG syntax because they introduce higher-order SVG primitives like texts, gradients or animations, also edits involving semantic understanding, multi-step reasoning (change many things at the same time), and design intent beyond what can be achieved through simple geometric or algorithmic transformations. These annotations involve operations such as adding new objects, integrating external SVG elements, inserting text with meaningful placement, restructuring layouts, or applying several coordinated edits simultaneously. They reflect realistic design actions performed by human experts and cannot be reproduced by rule-based procedures or low-level manipulations.

\subsection{VLM-as-Judge Metric for SVG Generation}\label{sec:vlmasajudge}

Traditional metrics for SVG generation fail to capture key aspects such as semantic correctness, structural validity, and instruction following in vector code. To address this, we introduce a task-specific \emph{VLM-as-a-Judge} (VLMAJ) evaluation protocol, validated via human correlation across all four VectorGym tasks. We summarize the full methodology, correlation analysis, and judge selection in Appendix~\ref{app:vlmaj}.

\subsection{Reinforcement Learning Method for Multi-Task SVG Training}\label{sec:RL}

We introduce a reinforcement learning method to train VLM models on VectorGym tasks. We fine-tune a \textbf{Qwen3-VL 8B Instruct} model using Reinforcement Learning from Rendering Feedback (RLRF)~\citep{rodriguez2025rendering} to jointly learn all four VectorGym tasks. For the \textit{Text-to-SVG}, \textit{SVG Editing}, and \textit{Sketch-to-SVG} tasks, the model outputs SVG code. To compute rewards, we render both the predicted and ground-truth SVGs into raster images and evaluate them using a combination of perceptual similarity metrics and pixel-space distances. For the \textit{SVG Captioning} task, where both the prediction and ground truth are textual descriptions of the SVG, the reward is defined as the embedding similarity between the two texts, using \texttt{BGE-M3} as the embedding model.

We train the 8B model on all four tasks simultaneously within a unified RL framework. Our optimization procedure primarily follows GRPO~\cite{shao2024deepseekmath}, with modifications inspired by~\cite{liu2025part}. Standard GRPO computes the advantage for each prompt by normalizing rewards \textit{within} the group of $K$ sampled responses. Given a prompt $x$ with reward set $\{ r_k \}_{k=1}^K$, the GRPO group-level advantage is
\begin{equation}
A^{\text{group}}_k
    = \frac{ r_k - \mathrm{mean}\left( \{ r_j \}_{j=1}^K \right) }
           { \mathrm{std}\left( \{ r_j \}_{j=1}^K \right) }.
\end{equation}

In contrast, our variant normalizes the centered rewards using the \textit{batch-level} standard deviation computed over all $N \times K$ samples in the mini-batch:
\begin{equation}
A^{\text{batch}}_i
    = \frac{ r_i - \mathrm{mean}\left( \{ r_j \}_{j=1}^K \right) }
           { \mathrm{std}\left( \{ r_j \}_{j=1}^{N \times K} \right) }.
\end{equation}
We use a rollout batch size of 168 samples per step. For each sample, the model generates 8 sampled rollouts, producing 1,344 rollouts per iteration. 
We train the model for 600 iterations on a single compute node with \mbox{8~$\times$~H200} GPUs, and the full run finishes in about two days (384 GPU-hours), improving the overall score from 58.74 to 66.05 (+7.31) across all four tasks. Inference cost remains unchanged.
We set the learning rate to $3\times10^{-6}$, the KL coefficient to $0.01$, and the sampling temperature to $1.0$. Each iteration performs exactly one policy update on its rollout batch, so neither gradient clipping nor PPO-style ratio clipping is ever triggered during optimization.

To improve training stability, we also apply curriculum learning. We treat the length of a response as a proxy for its difficulty and therefore sort the samples by response lengths. Because our dataset mixes four different tasks, we sort samples within each task according to response length and then draw tasks proportionally to their dataset frequencies to construct each minibatch. This strategy allows the model to progress from shorter and simpler examples toward longer and more complex ones, while maintaining task balance throughout training.

\subsection{Evaluation}
We describe the metrics used for evaluation in VectorGym, in addition to the VLM-as-Judge metric defined above.

\textbf{Visual Similarity.} For tasks that require visual reproduction (Sketch2SVG, Text2SVG), we measure similarity between generated and target SVGs after rendering them to pixels. We use pixel Mean Squared Error (MSE), perceptual similarity (LPIPS), and Dino, a deep feature metric that captures alignment in learned representations~\citep{oquab2023dinov2}.

\textbf{Semantic Accuracy.} For Text2SVG, we evaluate whether the generated SVG captures the intended semantic meaning of the text through CLIP-based similarity and the VLM-Judge metric. For SVG Editing, we rely exclusively on the VLM-Judge since CLIP does not align well with editing instructions or edited outputs.

\textbf{SVG Captioning Metrics.} For captioning, we report ROUGE-L F1 (0 to 100, higher is better), BGE-M3 cosine similarity (0 to 100, higher is better), and an LLM-based rubric score (GPT-5 mapped from 0 to 5 into 0 to 100). Metrics are computed pairwise over each reference and prediction caption, then averaged across the corpus.

\textbf{Human Evaluation.} A subset of outputs from the top performing models on the validation split is evaluated by expert annotators. They assess overall quality, semantic correctness, and task specific criteria (see Table~\ref{tab:averages}).

\textbf{Overall VectorGym Score.} We define an overall score for our benchmark, intended to measure multi-task performance across SVG generation from sketches and texts, complex editing of SVGs, and SVG understanding through captioning from code. 
First, we compute a task-specific score $S_{\text{task}}$ for each of the four tasks. For Sketch2SVG and SVG Editing, the score is the average of the VLM Judge, DINO, inverted MSE ($100 - \text{MSE}$), and inverted LPIPS ($100 - \text{LPIPS}$), ensuring all components contribute positively. For Text2SVG, we average the VLM Judge, CLIP, and DINO scores. For SVG Captioning, we average the VLM Judge, BGE-M3, and ROUGE scores. Finally, the overall VectorGym score is computed as the arithmetic mean of the four task-specific scores:
\begin{equation}
\text{VectorGym} = \frac{1}{4} \sum_{\tau \in \mathcal{T}} S_{\tau}
\end{equation}
where $\mathcal{T} = \{\text{Sketch}, \text{Edit}, \text{Text}, \text{Caption}\}$. All individual metrics are scaled to a range of [0, 100] prior to aggregation.

\section{Experiments}
\label{sec:experiments}

We conduct comprehensive evaluation across all four VectorGym tasks using state-of-the-art VLMs. Our experimental setup is designed to provide fair comparison while highlighting the unique challenges of SVG code generation.

\begin{figure*}[h]
    \centering
    \includegraphics[width=\linewidth]{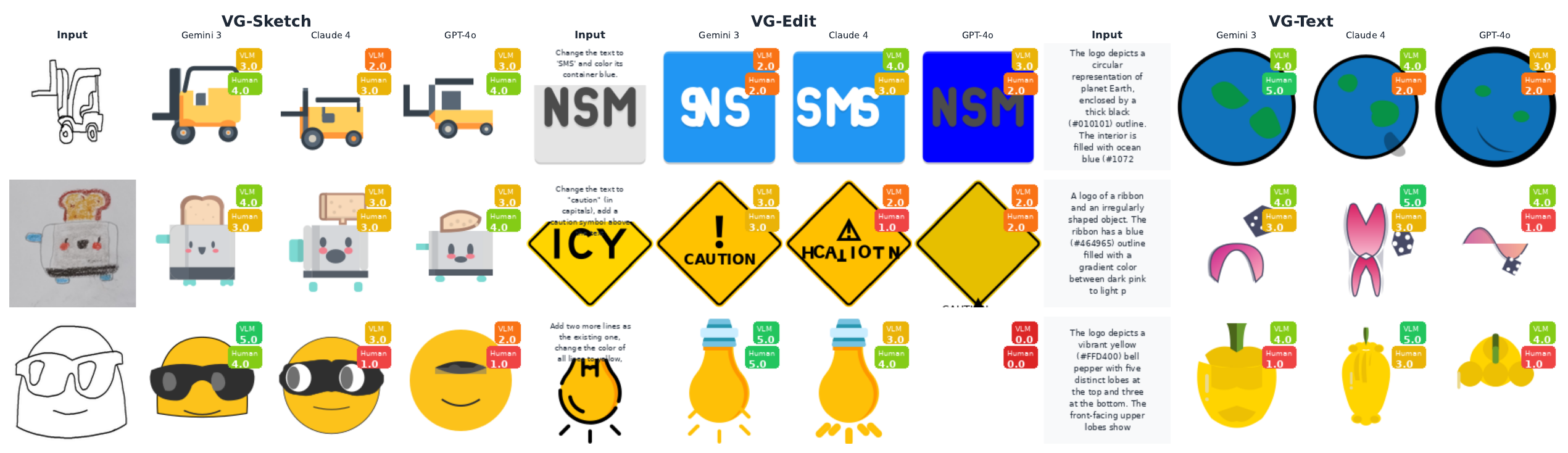}
\caption{\textbf{Qualitative results on VectorGym.} We display VLM-Judge and Human scores on a scale from 0 to 5. Each task shows three validation samples alongside the strongest models in our evaluation. Human ratings tend to be stricter, while VLM judges are more permissive and often cluster around mid-range values when uncertain.}
    \label{fig:human-vlm}
\end{figure*}

\subsection{Methods and Baselines}

We conduct a comprehensive evaluation using all available state-of-the-art VLMs that support code generation capabilities. Our baseline selection follows a systematic approach to ensure comprehensive coverage of the current landscape.

\textbf{In-Context Learning Experiments.} First, we evaluate the capabilities of frontier models on these tasks using in-context learning with a strong prompt to describe the task to perform. We include open and closed source models with the prompts specified in Appendix~\ref{sec:prompts-appendix}.

\textit{A. Closed-Source Models.} We evaluate leading commercial VLMs that demonstrate strong performance on visual understanding and code generation tasks: Gemini 2.5 Flash, Gemini 3 Pro, GPT-4o, GPT-5.1, and Claude Sonnet 4.5. These models represent the current state-of-the-art in multimodal understanding and have shown exceptional capabilities in various vision-language and code generation benchmarks.

\textit{B. Open-Source Models.} To ensure comprehensive coverage and reproducible research, we include leading open-source alternatives: Qwen2.5VL 32B-72B Instruct, Qwen3VL 8B-235B, and GLM4.5V 108B. We made best efforts to identify and include all available VLM models with public code implementations that could be executed on our tasks.

\textbf{RL Training Experiments.} As explained in Section~\ref{sec:RL}, we also train a Qwen3VL 8B Instruct model using the RLRF (Reinforcement Learning from Rendering Feedback) framework~\citep{rodriguez2025rendering}, which applies GRPO~\citep{shao2024deepseekmath} together with rendered SVG outputs to compute rewards. The model is trained on the VectorGym train split across all four tasks simultaneously.

\section{Results}
\label{sec:results}

We present a comprehensive evaluation of state-of-the-art VLMs across the four VectorGym tasks, highlighting capability gaps between proprietary and open-source models; qualitative VG-Sketch and VG-Edit examples are shown in \Cref{fig:vg-sketch-qualitative,fig:vg-edit-qualitative}.

\begin{table*}[ht]
\centering
\caption{\textbf{Sketch2SVG and SVG Editing Performance.} Metrics are reported such that higher values indicate better performance ($\uparrow$) or lower values indicate better performance ($\downarrow$). To compute the unified \textbf{Score}, MSE and LPIPS are inverted ($100 - x$) and averaged with VLM Judge and DINO, all scaled to $[0, 100]$. \textbf{Overall} represents the arithmetic mean of scores across all four tasks. The best results in each category are marked in \textbf{bold}.}
\label{tab:sketch_edit_perf}
\resizebox{\textwidth}{!}{%
\tabcolsep 2pt
\begin{tabular}{@{}lrrrrrrrrrrc@{}}
\toprule
 & \multicolumn{5}{c}{\textbf{Sketch2SVG}} & \multicolumn{5}{c}{\textbf{SVG Editing}} & \textbf{Overall} \\
\cmidrule(lr){2-6} \cmidrule(lr){7-11} \cmidrule(l){12-12}
\textbf{Model} & \textbf{VLM J} $\uparrow$ & \textbf{MSE} $\downarrow$ & \textbf{DINO} $\uparrow$ & \textbf{LPIPS} $\downarrow$ & \textbf{Score} $\uparrow$ & \textbf{VLM J} $\uparrow$ & \textbf{MSE} $\downarrow$ & \textbf{DINO} $\uparrow$ & \textbf{LPIPS} $\downarrow$ & \textbf{Score} $\uparrow$ & \textbf{VectorGym} $\uparrow$ \\
\midrule
\rowcolor{lightgray} \multicolumn{12}{@{}l}{\textbf{\textit{Open-source Models}}} \\
\midrule
Qwen2.5VL 72B Instruct     & 12.80 & 16.43 & 69.87 & 43.95 & 55.57 & 16.60 & 18.68 & 70.35 & 38.21 & 57.52 & \cellcolor{orange!20}44.27 \\
Qwen2.5VL 32B Instruct     & 17.80 & 15.15 & 71.63 & 42.65 & 57.91 & 20.20 & 17.04 & 72.31 & 37.05 & 59.61 & \cellcolor{orange!20}49.16 \\
GLM4.5V                    & 33.80 & 14.14 & 78.61 & 41.35 & 64.23 & 37.60 & 13.39 & 80.90 & 31.76 & 68.34 & \cellcolor{orange!20}57.02 \\
Qwen3VL 8B Instruct        & 33.00 & 13.76 & 81.01 & 40.97 & 64.82 & 57.40 & 11.01 & 90.44 & 25.27 & 77.89 & \cellcolor{orange!20}58.74 \\
Qwen3VL 235B Instruct      & 40.00 & 13.37 & 83.69 & 40.23 & 67.52 & 60.40 &  9.02 & 91.17 & 22.11 & 80.11 & \cellcolor{orange!20}62.32 \\
\midrule
Qwen3VL 8B Gym (Ours)      & \textbf{46.00} & \textbf{11.99} & \textbf{88.25} & \textbf{39.37} & \textbf{70.72} & \textbf{67.00} & \textbf{8.36} & \textbf{93.94} & \textbf{21.34} & \textbf{82.81} & \cellcolor{orange!20}\textbf{66.05} \\
\midrule
\rowcolor{lightgray} \multicolumn{12}{@{}l}{\textbf{\textit{Proprietary Models}}} \\
\midrule
Gemini 2.5 Flash           & 36.80 & 13.67 & 79.13 & 40.45 & 65.45 & 65.80 &  9.98 & 90.54 & 21.16 & 81.30 & \cellcolor{orange!20}61.42 \\
GPT-4o                     & 46.00 & 13.17 & 85.11 & 39.74 & 69.55 & 66.80 &  8.43 & 92.27 & 21.24 & 82.35 & \cellcolor{orange!20}64.93 \\
Claude Sonnet 4.5          & 58.80 & 12.54 & 88.42 & 39.29 & 73.85 & 79.40 &  6.29 & \textbf{95.61} & 16.46 & 88.07 & \cellcolor{orange!20}70.31 \\
GPT-5.1                    & 64.00 & 12.28 & 89.47 & 38.42 & 75.69 & 78.00 &  5.92 & 95.59 & 16.83 & 87.71 & \cellcolor{orange!20}71.36 \\
Gemini 3 Pro               & \textbf{72.20} & \textbf{11.31} & \textbf{89.78} & \textbf{36.43} & \textbf{78.56} & \textbf{81.20} & \textbf{5.89} & 95.55 & \textbf{16.01} & \textbf{88.71} & \cellcolor{orange!20}\textbf{73.17} \\
\bottomrule
\end{tabular}%
}
\end{table*}

\subsection{Sketch2SVG Generation}
The Sketch2SVG task evaluates the model's ability to infer vector geometry from raster sketches, a problem characterized by high ambiguity and visual abstraction. Figure~\ref{fig:human-vlm} (middle) shows results of this task among the best performing models, as well as human and VLMAJ scores. As shown in Table~\ref{tab:sketch_edit_perf}, \textbf{Gemini 3 Pro achieves the highest performance}, obtaining a Score of 78.56 and a VLM Judge score of 72.20. This indicates a superior capability in mapping pixel-level visual features to precise SVG path commands. GPT-5.1 follows with a Score of 75.69.

Notably, the performance gap between the top model and the open-source baseline is significant. However, \textbf{our proposed Qwen3VL 8B Gym model achieves a Score of 70.72, surpassing both GPT-4o (69.55) and the much larger Qwen3VL 235B (67.52)}. The Gym model's VLM Judge score (46.00) represents a substantial improvement over the base Qwen3VL 8B Instruct (33.00), validating the efficacy of curriculum learning for structural visual alignment.

\subsection{SVG Editing}
SVG Editing requires disjoint reasoning capabilities: parsing the existing code structure and manipulating it according to natural language instructions. \textbf{Gemini 3 Pro again leads this task with a Score of 88.71}, closely followed by Claude Sonnet 4.5 (88.07). Claude Sonnet 4.5 notably achieves the highest DINO score (95.61) , while Gemini 3 Pro achieves the lowest MSE (5.89), suggesting it generates visually faithful edits even if the structural implementation differs slightly from the ground truth.

\textbf{Our Qwen3VL 8B Gym demonstrates remarkable competitiveness in this domain}, achieving a Score of 82.81. This performance exceeds that of GPT-4o (82.35) and approaches the proprietary frontier. The low MSE (8.36) of the Gym model compared to the base 8B model (11.01) indicates that fine-tuning on edit trajectories significantly enhances the model's precision in coordinate manipulation.

\begin{table*}[t]
\centering
\caption{\textbf{Text2SVG and SVG Captioning Performance.} Higher values indicate better performance ($\uparrow$). DINO scores for Text2SVG are scaled to $[0, 100]$. The \textbf{Score} column represents the unweighted average of metrics within each task.}
\label{tab:text_cap_perf}
\resizebox{0.9\textwidth}{!}{%
\tabcolsep 2.5pt
\begin{tabular}{@{}lrrrrrrrr@{}}
\toprule
 & \multicolumn{4}{c}{\textbf{Text2SVG}} & \multicolumn{4}{c}{\textbf{SVG Captioning}} \\
\cmidrule(lr){2-5} \cmidrule(l){6-9}
\textbf{Model} & \textbf{VLM J} $\uparrow$ & \textbf{CLIP} $\uparrow$ & \textbf{DINO} $\uparrow$ & \textbf{Score} $\uparrow$ & \textbf{VLM J} $\uparrow$ & \textbf{BGE-M3} $\uparrow$ & \textbf{ROUGE} $\uparrow$ & \textbf{Score} $\uparrow$ \\
\midrule
\rowcolor{lightgray} \multicolumn{9}{@{}l}{\textbf{\textit{Open-source Models}}} \\
\midrule
Qwen2.5-VL-72B-Instruct     & 25.80   & 25.78 & 71.00 & 40.86 & 9.60  & 52.08 &  7.70 & 23.13 \\
Qwen3-VL-8B-Instruct        & 55.20   & 29.48 & 81.71 & 55.46 & 25.20 & 66.27 & 18.87 & 36.78 \\
GLM-4.5V                    & 59.40   & 28.91 & 80.44 & 56.25 & 38.00 & 62.85 & 16.86 & 39.24 \\
Qwen2.5-VL-32B-Instruct       & 22.60   & 24.95 & 68.96 & 38.84 & 38.40 & 66.10 & 16.35 & 40.28 \\
Qwen3-VL-235B-Instruct      & 66.80   & 29.60 & 82.63 & 59.68 & \textbf{40.40} & 67.14 & 18.33 & 41.96 \\
Qwen3-VL-8B-Gym (Ours)      & \textbf{72.80}  & \textbf{30.55} & \textbf{87.46} & \textbf{63.60} & 35.80 & \textbf{79.76} & \textbf{25.58} & \textbf{47.05} \\
\midrule
\rowcolor{lightgray} \multicolumn{9}{@{}l}{\textbf{\textit{Proprietary Models}}} \\
\midrule
GPT-4o                      & 74.60   & 30.43 & 84.23 & 63.09 & 46.00 & 66.82 & 21.33 & 44.72 \\
Gemini 2.5 Flash            & 54.00   & 27.67 & 77.65 & 53.11 & 45.80 & 69.24 & 22.45 & 45.83 \\
Claude Sonnet 4.5           & 89.00   & \textbf{30.91} & 87.66 & 69.19 & 59.20 & 70.17 & 21.08 & 50.15 \\
GPT-5.1                     & \textbf{93.00}   & 30.83 & 88.20 & \textbf{70.68} & 62.20 & 70.45 & 21.49 & 51.38 \\
Gemini 3 Pro                & 89.80   & 30.87 & \textbf{89.09} & 69.92 & \textbf{70.40} & \textbf{72.27} & \textbf{23.83} & \textbf{55.50} \\
\bottomrule
\end{tabular}%
}
\end{table*}

\subsection{Text2SVG Generation}
Table~\ref{tab:text_cap_perf} presents our Text2SVG generation results, revealing clear performance hierarchies and interesting patterns. 
Among proprietary models, \textbf{GPT-5.1 achieves state-of-the-art performance} with a VLM Judge score of 93.00 and an overall Score of 70.68. The proprietary models exhibit a relatively narrow performance band, with Gemini 3 Pro (69.92) and Claude Sonnet 4.5 (69.19) performing comparably.

Among open-source models, \textbf{our fine-tuned Qwen3VL 8B Gym outperforms the larger Qwen3VL 235B baseline} (Score: 63.60 vs. 59.68) and achieves parity with GPT-4o (63.09). This result emphasizes that for well-defined generation tasks, specialized smaller models can effectively compete with general-purpose frontier models.

\subsection{SVG Captioning}

The SVG Captioning results in Table~\ref{tab:text_cap_perf} reveal interesting patterns distinct from the generation tasks. \textbf{Gemini 3 Pro dominates the VLMAJ metric (70.40)}, significantly outperforming other models, which aligns with its ability to map code structure back to high-level semantic descriptions. However, the traditional NLP metrics show different rankings: our Qwen3VL 8B Gym achieves the highest BGE-M3 (79.76) and ROUGE scores (25.58) across the entire benchmark. \textbf{Qwen3VL 8B Gym outperforms all proprietary models in keyword-based metrics.} This discrepancy between its state-of-the-art retrieval scores and its lower VLMAJ score (35.80 compared to 40.40 for the Qwen3VL 235B baseline) suggests that while the Gym model captures salient semantic details, it may lack the conversational fluency or formatting preference favored by the VLM Judge.

\subsection{Cross-Task Analysis}

Our comprehensive evaluation across Text2SVG, SVG Editing, and Sketch2SVG reveals several critical insights about current VLM capabilities in vector graphics generation.

\textbf{Overall Performance Hierarchy.} Aggregating across all tasks, Gemini 3 Pro achieves the highest VectorGym score of 73.17, followed by GPT-5.1 (71.36). This establishes Gemini 3 Pro as the most capable model for multimodal code-visual reasoning tasks.

\textbf{Effectiveness of RL.} The Qwen3VL 8B Gym model achieves an overall score of 66.05, surpassing GPT-4o (64.93) and substantially outperforming its larger counterpart, Qwen3VL 235B (62.32). This finding validates the hypothesis that the limitations of smaller parameter counts can be effectively offset by high-quality, task-specific curriculum learning in the SVG domain.

\textbf{Task Complexity.} The results establish a clear difficulty hierarchy: Text2SVG (easiest, GPT-5.1: 93.00) $>$ SVG Editing (intermediate, Gemini 3 Pro: 81.20) $>$ Sketch2SVG (Gemini 3 Pro: 72.20) $>$ SVG Captioning (hardest, Gemini 3 Pro: 70.40). This ranking aligns with intuitive expectations: text descriptions provide explicit semantic guidance, editing requires understanding existing structures, sketches demand interpretation of imprecise visual input, while captioning requires the rigorous abstraction of high-level semantics from low-level geometric code. 
VLMs perform substantially better with explicit text than with sketches: GPT-5.1 scores 93.00 on Text2SVG, whereas Gemini 3 Pro reaches 72.20 on Sketch2SVG; open-source Qwen3VL 235B and 8B score 40.00 and 33.00. RL raises the 8B score to 46.00, but the gap to closed models persists, indicating a limitation beyond data or model scale. Models also rarely add color absent from the input sketch (Figure~\ref{fig:sketch-app}).

\section{Conclusion}
\label{sec:conclusion}

We introduced VectorGym, a comprehensive multi-task benchmark for SVG code generation and manipulation encompassing Sketch2SVG, SVG editing, Text2SVG, and SVG captioning. VectorGym introduces the novel Sketch2SVG task and releases the first dataset of complex, human-authored SVG edits, with gold-standard human annotations across all tasks. Beyond the benchmark, we provide a multi-task reinforcement learning baseline using rendering-based rewards that trains a Qwen3-VL 8B model to achieve state-of-the-art performance among open-source models, surpassing much larger models and matching GPT-4o. Our evaluation of frontier VLMs reveals significant performance gaps, with Gemini 3 Pro leading overall. VectorGym establishes a rigorous evaluation standard for visual code generation and provides a framework for advancing SVG generation capabilities.


\paragraph{Limitations}
\label{sec:limitations}

VectorGym expands the range of capabilities that can be evaluated and optimized for fine grained control of state of the art SVG models. We tested several leading models in a zero shot setting, and we also ran RL training experiments that produced strong results. 
Our held-out split measures generalization within SVG-Stack, not transfer to other corpora or design-tool exports. Future work should extend dataset curation and evaluation across sources and improve training for sketch generation and complex editing.

\paragraph{Impact Statement}
This paper introduces a benchmark for evaluating SVG generation and manipulation capabilities in vision-language models. As a benchmark contribution, our work primarily impacts the research community by enabling more rigorous evaluation of visual code generation systems. We do not foresee direct negative societal consequences from the benchmark itself, though we acknowledge that advances in SVG generation could eventually impact professional design workflows. We have addressed potential representation biases through careful data curation and filtering, as noted in our Ethics Statement.

\paragraph{Ethics Statement}
\label{sec:ethics}

The models evaluated in this benchmark may exhibit biases inherited from their training data, potentially affecting the fairness and representation of generated SVG content across different demographics, cultures, and artistic styles. We have performed extensive filtering and human curation to ensure VectorGym does not include such instances.


\bibliography{main}

\begin{thebibliography}{36}
\providecommand{\natexlab}[1]{#1}

\bibitem[{Bai et~al.(2025{\natexlab{a}})Bai, Cai, Chen, Chen, Chen, Cheng,
  Deng, Ding, Gao, Ge et~al.}]{bai2025qwen3vl}
Shuai Bai, Yuxuan Cai, Ruizhe Chen, Keqin Chen, Xionghui Chen, Zesen Cheng,
  Lianghao Deng, Wei Ding, Chang Gao, Chunjiang Ge, and 1 others.
  2025{\natexlab{a}}.
\newblock {Qwen3-VL} technical report.
\newblock \emph{arXiv preprint arXiv:2511.21631}.

\bibitem[{Bai et~al.(2025{\natexlab{b}})Bai, Chen, Liu, Wang, Ge, Song, Dang,
  Wang, Wang, Tang et~al.}]{bai2025qwen25vl}
Shuai Bai, Keqin Chen, Xuejing Liu, Jialin Wang, Wenbin Ge, Sibo Song, Kai
  Dang, Peng Wang, Shijie Wang, Jun Tang, and 1 others. 2025{\natexlab{b}}.
\newblock {Qwen2.5-VL} technical report.
\newblock \emph{arXiv preprint arXiv:2502.13923}.

\bibitem[{Belouadi et~al.(2023)Belouadi, Lauscher, and
  Eger}]{belouadi2023automatikz}
Jonas Belouadi, Anne Lauscher, and Steffen Eger. 2023.
\newblock Automatikz: Text-guided synthesis of scientific vector graphics with
  tikz.
\newblock \emph{arXiv preprint arXiv:2310.00367}.

\bibitem[{Cai et~al.(2023)Cai, Huang, Li, Wang, and Lee}]{cai2023leveraging}
Mu~Cai, Zeyi Huang, Yuheng Li, Haohan Wang, and Yong~Jae Lee. 2023.
\newblock Leveraging large language models for scalable vector graphics-driven
  image understanding.
\newblock \emph{arXiv preprint arXiv:2306.06094}.

\bibitem[{Cao et~al.(2023)Cao, Wang, Echevarria, and Liu}]{Cao2023SVGformer}
Defu Cao, Zhaowen Wang, Jose Echevarria, and Yan Liu. 2023.
\newblock Svgformer: Representation learning for continuous vector graphics
  using transformers.
\newblock In \emph{Proceedings of the IEEE/CVF Conference on Computer Vision
  and Pattern Recognition}, pages 10093--10102.

\bibitem[{Carlier et~al.(2020)Carlier, Danelljan, Alahi, and
  Timofte}]{Carlier2020DeepSVG}
Alexandre Carlier, Martin Danelljan, Alexandre Alahi, and Radu Timofte. 2020.
\newblock \href
  {https://proceedings.neurips.cc/paper/2020/file/bcf9d6bd14a2095866ce8c950b702341-Paper.pdf}
  {Deepsvg: A hierarchical generative network for vector graphics animation}.
\newblock In \emph{NeurIPS}.

\bibitem[{Chen et~al.(2025)Chen, Dong, Xu, Wu, Tang, Zhang, Yan, Wu, Zhang, Hou
  et~al.}]{chen2025svgenius}
Siqi Chen, Xinyu Dong, Haolei Xu, Xingyu Wu, Fei Tang, Hang Zhang, Yuchen Yan,
  Linjuan Wu, Wenqi Zhang, Guiyang Hou, and 1 others. 2025.
\newblock Svgenius: Benchmarking llms in svg understanding, editing and
  generation.
\newblock In \emph{Proceedings of the 33rd ACM International Conference on
  Multimedia}, pages 13289--13296.

\bibitem[{Clou\^{a}tre and Demers(2019)}]{Clouatre2019FIGR8}
Louis Clou\^{a}tre and Marc Demers. 2019.
\newblock \href {https://arxiv.org/abs/1901.02199} {Figr: Few-shot image
  generation with reptile}.
\newblock \emph{arXiv:1901.02199}.

\bibitem[{Comanici et~al.(2025)Comanici, Bieber, Schaekermann, Pasupat,
  Sachdeva, Dhillon, Blistein, Ram, Zhang, Rosen et~al.}]{comanici2025gemini}
Gheorghe Comanici, Eric Bieber, Mike Schaekermann, Ice Pasupat, Noveen
  Sachdeva, Inderjit Dhillon, Marcel Blistein, Ori Ram, Dan Zhang, Evan Rosen,
  and 1 others. 2025.
\newblock Gemini 2.5: Pushing the frontier with advanced reasoning,
  multimodality, long context, and next generation agentic capabilities.
\newblock \emph{arXiv preprint arXiv:2507.06261}.

\bibitem[{Ferraiolo et~al.(2000)Ferraiolo, Jun, and
  Jackson}]{ferraiolo2000scalable}
Jon Ferraiolo, Fujisawa Jun, and Dean Jackson. 2000.
\newblock \emph{Scalable vector graphics (SVG) 1.0 specification}.
\newblock iuniverse Bloomington.

\bibitem[{Jain et~al.(2023)Jain, Xie, and Abbeel}]{jain2023vectorfusion}
Ajay Jain, Amber Xie, and Pieter Abbeel. 2023.
\newblock Vectorfusion: Text-to-svg by abstracting pixel-based diffusion
  models.
\newblock In \emph{Proceedings of the IEEE/CVF Conference on Computer Vision
  and Pattern Recognition}, pages 1911--1920.

\bibitem[{Kocetkov et~al.(2022)Kocetkov, Li, Allal, Li, Mou, Ferrandis,
  Jernite, Mitchell, Hughes, Wolf et~al.}]{kocetkov2022stack}
Denis Kocetkov, Raymond Li, Loubna~Ben Allal, Jia Li, Chenghao Mou,
  Carlos~Mu{\~n}oz Ferrandis, Yacine Jernite, Margaret Mitchell, Sean Hughes,
  Thomas Wolf, and 1 others. 2022.
\newblock The stack: 3 tb of permissively licensed source code.
\newblock \emph{arXiv preprint arXiv:2211.15533}.

\bibitem[{Li et~al.(2025)Li, Yu, Wei, Dong, Lin, Yang, Wang, and
  Hao}]{li2025unisvg}
Jinke Li, Jiarui Yu, Chenxing Wei, Hande Dong, Qiang Lin, Liangjing Yang,
  Zhicai Wang, and Yanbin Hao. 2025.
\newblock Unisvg: A unified dataset for vector graphic understanding and
  generation with multimodal large language models.
\newblock \emph{arXiv preprint arXiv:2508.07766}.

\bibitem[{Li et~al.(2020)Li, Luk{\'a}{\v{c}}, Gharbi, and
  Ragan-Kelley}]{DiffVG2020}
Tzu-Mao Li, Michal Luk{\'a}{\v{c}}, Micha{\"e}l Gharbi, and Jonathan
  Ragan-Kelley. 2020.
\newblock \href {https://people.csail.mit.edu/tzumao/diffvg/} {Differentiable
  vector graphics rasterization for editing and learning}.
\newblock \emph{ACM TOG (SIGGRAPH Asia)}.

\bibitem[{Liu et~al.(2025)Liu, Liu, He, Wang, Liu, Pan, Hu, Xiong, Huang, Hu
  et~al.}]{liu2025part}
Zihe Liu, Jiashun Liu, Yancheng He, Weixun Wang, Jiaheng Liu, Ling Pan, Xinyu
  Hu, Shaopan Xiong, Ju~Huang, Jian Hu, and 1 others. 2025.
\newblock Part i: Tricks or traps? a deep dive into rl for llm reasoning.
\newblock \emph{arXiv preprint arXiv:2508.08221}.

\bibitem[{Lopes et~al.(2019)Lopes, Ha, Eck, and Shlens}]{Lopes2019SVGFonts}
Raphael~Gontijo Lopes, David Ha, Douglas Eck, and Jonathon Shlens. 2019.
\newblock A learned representation for scalable vector graphics.
\newblock In \emph{Proceedings of the IEEE/CVF International Conference on
  Computer Vision}, pages 7930--7939.

\bibitem[{Ma{\~n}as et~al.(2024)Ma{\~n}as, Krojer, and
  Agrawal}]{manas2024improving}
Oscar Ma{\~n}as, Benno Krojer, and Aishwarya Agrawal. 2024.
\newblock Improving automatic vqa evaluation using large language models.
\newblock In \emph{Proceedings of the AAAI Conference on Artificial
  Intelligence}, volume~38, pages 4171--4179.

\bibitem[{Nishina and Matsui(2024)}]{nishina2024svgeditbench}
Kunato Nishina and Yusuke Matsui. 2024.
\newblock Svgeditbench: A benchmark dataset for quantitative assessment of
  llm's svg editing capabilities.
\newblock In \emph{Proceedings of the IEEE/CVF Conference on Computer Vision
  and Pattern Recognition}, pages 8142--8147.

\bibitem[{Nishina and Matsui(2025)}]{nishina2025svgeditbench}
Kunato Nishina and Yusuke Matsui. 2025.
\newblock Svgeditbench v2: A benchmark for instruction-based svg editing.
\newblock \emph{arXiv preprint arXiv:2502.19453}.

\bibitem[{OpenAI(2023)}]{openai2023gpt4}
OpenAI. 2023.
\newblock \href {https://arxiv.org/abs/2303.08774} {Gpt-4 technical report}.
\newblock \emph{Preprint}, arXiv:2303.08774.

\bibitem[{Oquab et~al.(2023)Oquab, Darcet, Moutakanni, Vo, Szafraniec,
  Khalidov, Fernandez, Haziza, Massa, El-Nouby et~al.}]{oquab2023dinov2}
Maxime Oquab, Timoth{\'e}e Darcet, Th{\'e}o Moutakanni, Huy Vo, Marc
  Szafraniec, Vasil Khalidov, Pierre Fernandez, Daniel Haziza, Francisco Massa,
  Alaaeldin El-Nouby, and 1 others. 2023.
\newblock Dinov2: Learning robust visual features without supervision.
\newblock \emph{arXiv preprint arXiv:2304.07193}.

\bibitem[{Quint(2003)}]{quint2003scalable}
Antoine Quint. 2003.
\newblock Scalable vector graphics.
\newblock \emph{IEEE MultiMedia}, 10(3):99--102.

\bibitem[{Rodriguez et~al.(2025{\natexlab{a}})Rodriguez, Jian, Panigrahi,
  Zhang, Feizi, Puri et~al.}]{rodriguez2025bigdocs}
Juan~A. Rodriguez, Xiangru Jian, Siba~Smarak Panigrahi, Tianyu Zhang, Aarash
  Feizi, Abhay Puri, and 1 others. 2025{\natexlab{a}}.
\newblock {BigDocs}: An open dataset for training multi-modal models on
  document and code tasks.
\newblock In \emph{International Conference on Learning Representations}.

\bibitem[{Rodriguez et~al.(2025{\natexlab{b}})Rodriguez, Puri, Agarwal,
  Laradji, Rodriguez, Rajeswar, Vazquez, Pal, and
  Pedersoli}]{Rodriguez2025StarVector}
Juan~A Rodriguez, Abhay Puri, Shubham Agarwal, Issam~H Laradji, Pau Rodriguez,
  Sai Rajeswar, David Vazquez, Christopher Pal, and Marco Pedersoli.
  2025{\natexlab{b}}.
\newblock Starvector: Generating scalable vector graphics code from images and
  text.
\newblock In \emph{Proceedings of the Computer Vision and Pattern Recognition
  Conference}, pages 16175--16186.

\bibitem[{Rodriguez et~al.(2023{\natexlab{a}})Rodriguez, Vazquez, Laradji,
  Pedersoli, and Rodriguez}]{rodriguez2023figgen}
Juan~A Rodriguez, David Vazquez, Issam Laradji, Marco Pedersoli, and Pau
  Rodriguez. 2023{\natexlab{a}}.
\newblock Figgen: Text to scientific figure generation.
\newblock \emph{arXiv preprint arXiv:2306.00800}.

\bibitem[{Rodriguez et~al.(2023{\natexlab{b}})Rodriguez, Vazquez, Laradji,
  Pedersoli, and Rodriguez}]{rodriguez2023ocr}
Juan~A Rodriguez, David Vazquez, Issam Laradji, Marco Pedersoli, and Pau
  Rodriguez. 2023{\natexlab{b}}.
\newblock Ocr-vqgan: Taming text-within-image generation.
\newblock In \emph{Proceedings of the IEEE/CVF Winter Conference on
  Applications of Computer Vision}, pages 3689--3698.

\bibitem[{Rodriguez et~al.(2025{\natexlab{c}})Rodriguez, Zhang, Puri, Feizi,
  Pramanik, Wichmann, Mondal, Samsami, Awal, Taslakian
  et~al.}]{rodriguez2025rendering}
Juan~A Rodriguez, Haotian Zhang, Abhay Puri, Aarash Feizi, Rishav Pramanik,
  Pascal Wichmann, Arnab Mondal, Mohammad~Reza Samsami, Rabiul Awal, Perouz
  Taslakian, and 1 others. 2025{\natexlab{c}}.
\newblock Rendering-aware reinforcement learning for vector graphics
  generation.
\newblock \emph{arXiv preprint arXiv:2505.20793}.

\bibitem[{Rombach et~al.(2021)Rombach, Blattmann, Lorenz, Esser, and
  Ommer}]{rombach2021highresolution}
Robin Rombach, Andreas Blattmann, Dominik Lorenz, Patrick Esser, and Björn
  Ommer. 2021.
\newblock \href {https://arxiv.org/abs/2112.10752} {High-resolution image
  synthesis with latent diffusion models}.
\newblock \emph{Preprint}, arXiv:2112.10752.

\bibitem[{Shao et~al.(2024)Shao, Wang, Zhu, Xu, Song, Bi, Zhang, Zhang, Li, Wu
  et~al.}]{shao2024deepseekmath}
Zhihong Shao, Peiyi Wang, Qihao Zhu, Runxin Xu, Junxiao Song, Xiao Bi, Haowei
  Zhang, Mingchuan Zhang, YK~Li, Yang Wu, and 1 others. 2024.
\newblock Deepseekmath: Pushing the limits of mathematical reasoning in open
  language models.
\newblock \emph{arXiv preprint arXiv:2402.03300}.

\bibitem[{Vinker et~al.(2022)Vinker, Pajouheshgar, Bo, Bachmann, Bermano,
  Cohen-Or, Zamir, and Shamir}]{vinker2022clipasso}
Yael Vinker, Ehsan Pajouheshgar, Jessica~Y Bo, Roman~Christian Bachmann,
  Amit~Haim Bermano, Daniel Cohen-Or, Amir Zamir, and Ariel Shamir. 2022.
\newblock Clipasso: Semantically-aware object sketching.
\newblock \emph{ACM Transactions on Graphics (TOG)}, 41(4):1--11.

\bibitem[{{Vision Cortex}(2023)}]{vtracer}
{Vision Cortex}. 2023.
\newblock {VTracer}.
\newblock \url{https://www.visioncortex.org/vtracer-docs}.

\bibitem[{Wu et~al.(2023)Wu, Su, Ma, and Liao}]{wu2023iconshop}
Ronghuan Wu, Wanchao Su, Kede Ma, and Jing Liao. 2023.
\newblock Iconshop: Text-based vector icon synthesis with autoregressive
  transformers.
\newblock \emph{arXiv preprint arXiv:2304.14400}.

\bibitem[{Xing et~al.(2025)Xing, Hu, Liang, Zhang, Xu, and
  Yu}]{xing2025empowering}
Ximing Xing, Juncheng Hu, Guotao Liang, Jing Zhang, Dong Xu, and Qian Yu. 2025.
\newblock Empowering llms to understand and generate complex vector graphics.
\newblock In \emph{Proceedings of the Computer Vision and Pattern Recognition
  Conference}, pages 19487--19497.

\bibitem[{Yang et~al.(2025)Yang, Cheng, Chen, Zeng, Yin, Zhang, Wang, Yu, Ma,
  and Jiang}]{yang2025omnisvg}
Yiying Yang, Wei Cheng, Sijin Chen, Xianfang Zeng, Fukun Yin, Jiaxu Zhang, Liao
  Wang, Gang Yu, Xingjun Ma, and Yu-Gang Jiang. 2025.
\newblock Omnisvg: A unified scalable vector graphics generation model.
\newblock \emph{arXiv preprint arXiv:2504.06263}.

\bibitem[{Zhang et~al.(2023)Zhang, Liu, Zhang, Cheng, and
  Wang}]{zhang2023pixelsexploringhumanreadablesvg}
Tong Zhang, Haoyang Liu, Peiyan Zhang, Yuxuan Cheng, and Haohan Wang. 2023.
\newblock \href {https://arxiv.org/abs/2311.15543} {Beyond pixels: Exploring
  human-readable svg generation for simple images with vision language models}.
\newblock \emph{Preprint}, arXiv:2311.15543.

\bibitem[{Zou et~al.(2024)Zou, Cai, Zhang, and Lee}]{VGBench2024}
Bocheng Zou, Mu~Cai, Jianrui Zhang, and Yong~Jae Lee. 2024.
\newblock Vgbench: Evaluating large language models on vector graphics
  understanding and generation.
\newblock \emph{arXiv preprint arXiv:2407.10972}.

\end{thebibliography}

\newpage
\appendix
\section{VectorGym Data Creation}
\label{svg-datasets-appendix}
\begin{figure*}
  \centering
  \includegraphics[width=1.0\linewidth]{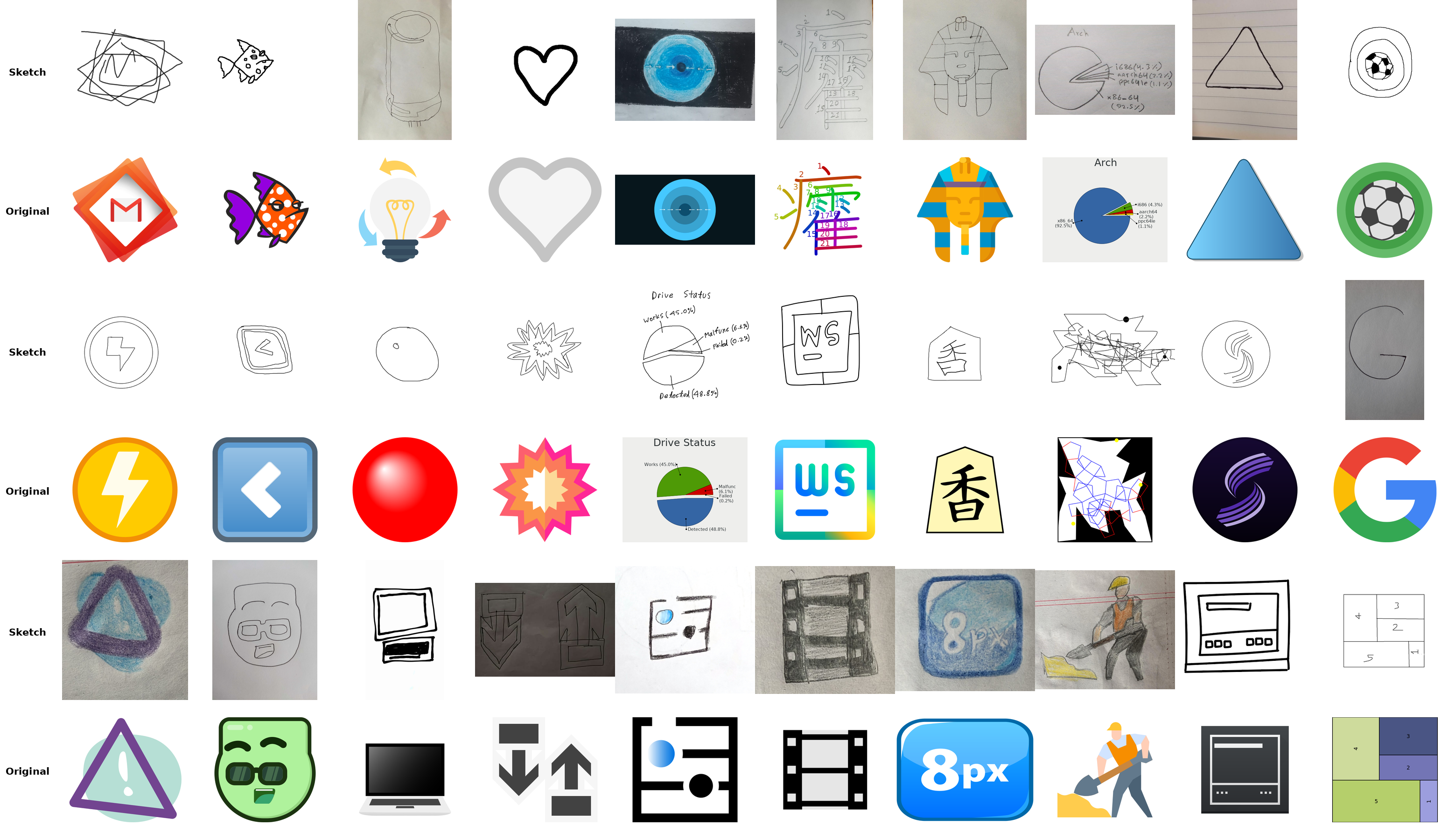}
  \caption{\textbf{Visualization of VG-Sketch Test Examples.} We randomly sample 30 examples, and show the sketch and the target vector.}
  \label{fig:vg-sketch-samples}
\end{figure*}

Here we provide additional details on the VectorGym datasets. Figures~\ref{fig:vg-edit-samples} and \ref{fig:vg-sketch-samples} illustrate test samples for the Sketch2SVG (VG-Sketch) and SVG Editing (VG-Edit) tasks. We further describe the annotation methodology, data creation and sampling process, annotation details, and task definitions. 

\subsection{Annotation Methodology}
\label{sec:annotation-details}

\begin{table}[t]
\centering
\caption{\textbf{VectorGym Curation Summary.} Dataset scale, annotation cost, and quality controls.}
\label{tab:curation_summary}
\small
\setlength{\tabcolsep}{4pt}
\renewcommand{\arraystretch}{1.05}
\begin{tabular}{@{}p{0.43\columnwidth}p{0.49\columnwidth}@{}}
\toprule
\textbf{Statistic} & \textbf{Value} \\
\midrule
Token-length filter &
2k--8k tokens \\
Color-entropy threshold &
$>0.55$ \\
Annotated SVGs &
6.9k \\
Annotation workforce &
2 vendors; 20+ annotators \\
Quality control &
Syntax/render checks; expert review \\
Annotation cost &
Approx. \$45k total; \$6.5/SVG \\
Ground-truth rating &
4.4--4.8/5 \\
Remaining source pool &
Approx. 2.3M SVGs \\
\bottomrule
\end{tabular}
\end{table}

The pipeline is format-agnostic except for the SVG renderer and complexity heuristics; annotation, quality control, and VLM-judge calibration (Appendix~\ref{app:vlmaj}) transfer directly to other visual-code formats.

\subsubsection{Data Curation and Sampling}

We extracted 7,000 high-quality samples from the SVG-Stack dataset through a rigorous multi-stage filtering process:

\textbf{Visual Quality Assessment:} Human experts manually reviewed SVG samples to identify visually appealing and well-formed graphics, filtering out corrupted, overly simplistic, or poorly designed samples.

\textbf{Token Length Filtering:} We applied token length constraints (2,000-8,000 tokens) to ensure meaningful complexity while maintaining computational feasibility. This range captures rich, detailed SVGs without exceeding practical processing limits for current VLMs.

\textbf{Color Entropy Thresholding:} We computed color entropy for each SVG to ensure visual diversity, filtering samples with insufficient color variation or monotonic palettes.

\textbf{Random Sampling:} Final samples were randomly selected to avoid systematic biases in content distribution. 

From the curated set of 7,000 samples, we kept the 300 items that originally belonged to the SVG Stack test split as our test set to avoid any train and test contamination. We also selected 100 samples from the training split for validation, which we used during development for method tuning, and for the human evaluation and correlation study used to design our VLM as a judge metric (see Appendix~\ref{app:vlmaj}).


\subsubsection{Annotation Vendor Partnership}

We partnered with two specialized data annotation vendors to ensure task-specific expertise:

\textbf{Vendor 1 - Sketch and Caption Generation:} Specialized in visual content creation, responsible for sketch generation and text descriptions. Annotators were equipped with professional drawing tools (digital tablets, cameras for hand-drawn sketches) and trained on SVG visual analysis.

\textbf{Vendor 2 - SVG Editing:} Focused on technical SVG manipulation, staffed with annotators having design and vector graphics backgrounds. We developed custom SVG editing tools specifically for this project to enable precise modifications.

\subsubsection{Annotator Demographics and Training}

Our annotation team comprised over 20 annotators with diverse demographics and gender representation. All annotators underwent specialized training:

\textbf{Technical Requirements:} Background in design, vector graphics, or coding. Annotators were tested on SVG understanding and tool proficiency before assignment.

\textbf{Equipment and Tools:} Professional cameras for photographing hand-drawn sketches, digital drawing tablets, custom SVG editing software, and standardized annotation interfaces.

\subsubsection{Task-Specific Annotation Procedures}

\textbf{Sketch2SVG Generation:} Annotators were provided with SVG images and asked to create corresponding sketches in two variants:
\begin{itemize}
    \item \textbf{Hand-drawn:} Using pen or pencil on paper, photographed with standardized lighting and resolution
    \item \textbf{Digital:} Created using drawing tablets and stylus input for consistent digital sketches
\end{itemize}
Both variants included colored and black-and-white versions to test model robustness across different input modalities.

\textbf{SVG Editing - Ensuring Complexity:} We implemented strict complexity requirements to avoid trivial edits that could be synthetically generated:

\textit{Prohibited Simple Edits:} Rotation, color changes, scaling, basic shape removal - operations easily automated by current LLMs.

\textit{Required Complex Edits:} Path modifications, primitive additions, parameter adjustments, conceptual additions requiring semantic understanding. For example:
\begin{itemize}
    \item Adding elements from other SVGs in the database (e.g., incorporating a needle shape into a hammer SVG)
    \item Modifying facial expressions in character illustrations
    \item Converting chart types (pie to bar charts)
    \item Structural modifications requiring new geometric primitives
\end{itemize}

\textbf{Caption Generation:} We implemented a comprehensive multi-stage process for generating high-quality text descriptions:

\begin{enumerate}
    \item \textbf{Detailed Visual Description:} Annotators created comprehensive descriptions of vector graphics, with particular emphasis on color specification. To ensure color accuracy, annotators were required to include hexadecimal color codes in parentheses alongside natural language color descriptions (e.g., "red (\#FF0000)").
    
    \item \textbf{Cross-validation with VLM:} All human-generated descriptions were processed and cross-validated using Qwen2-VL-32B to ensure consistency and completeness of visual descriptions.
    
    \item \textbf{Instruction Reformatting:} Captions were systematically reformatted from descriptive statements into instruction-style prompts suitable for the Text2SVG generation task. This process generated two distinct variants:
    \begin{itemize}
        \item \textbf{Hexadecimal Color Version:} Instructions containing precise hexadecimal color specifications, which empirically demonstrate superior SVG generation accuracy
        \item \textbf{Natural Language Color Version:} Instructions using standard color names for broader accessibility
    \end{itemize}
    
    \item \textbf{Quality Validation:} Final consistency checks and inter-annotator agreement measurement across all caption variants
\end{enumerate}

\textbf{Quality Assurance:} All annotations underwent rigorous quality control including automated SVG syntax validation, human verification of task requirements, and consistency checks across related task pairs.

\subsection{VLM-as-Judge Evaluation Metric for SVG Generation}\label{app:vlmaj}

Traditional evaluation metrics for SVG generation (typically based on image reconstruction or text–image alignment) often fall short in capturing the nuanced visual and semantic qualities that determine the success of generated vector graphics~\citep{Rodriguez2025StarVector,li2025unisvg, chen2025svgenius}. Existing work lacks comprehensive evaluation frameworks tailored to SVG generation, particularly metrics that can jointly assess visual fidelity and semantic alignment in vector code outputs~\citep{VGBench2024, nishina2025svgeditbench}.

\begin{table*}[t]
\centering
\small
\caption{\textbf{VLM as a Judge and Human Correlation Analysis.} We run generation on the tasks for Claude 4.5, Gemini 3 Pro, and GPT-4o, and evaluate outputs using a range of VLMs (both closed and open, large models) to score them with the prompts presented. We also collect human ratings using the same instructions given to VLM judges, then compute Pearson correlation to identify the best VLMs as judges. The evaluation uses 100 validation samples extracted from the training set. Results show Gemini 3 Pro is generally the best judge, except for SVG Editing, where GPT-5.1 achieves the highest average correlation with human judgments (0.70). Sketch and text tasks show lower correlations, likely due to the more creative nature of these tasks.}
\label{tab:vlm_correlation}
\resizebox{\textwidth}{!}{
\begin{tabular}{@{}llccccccc@{}}
\toprule
\textbf{Task} & \textbf{Generator} & \multicolumn{7}{c}{\textbf{Models used as Judges}} \\
\cmidrule(lr){3-9}
 & & Claude 4.5 Sonnet & Gemini 2.5 Flash & Gemini 3 Pro & GPT 5.1 & Qwen2.5VL 72B & Qwen3.VL 235B & GLM4.5 355B \\
\midrule
\multirow{5}{*}{\rotatebox[origin=c]{90}{\textbf{VG-Sketch}}} & \cellcolor{gray!15}Ground Truth & \cellcolor{gray!15}1.00 & \cellcolor{gray!15}1.00 & \cellcolor{gray!15}1.00 & \cellcolor{gray!15}1.00 & \cellcolor{gray!15}1.00 & \cellcolor{gray!15}1.00 & \cellcolor{gray!15}-0.07 \\
& Claude 4.5 Sonnet & \cellcolor{green!20}0.63 & \cellcolor{green!20}0.73 & \cellcolor{green!20}0.72 & \cellcolor{green!20}0.62 & \cellcolor{green!20}0.57 & \cellcolor{green!20}0.69 & \cellcolor{green!20}0.67 \\
& Gemini 3 Pro & \cellcolor{green!20}0.79 & \cellcolor{green!20}0.82 & \cellcolor{green!20}0.80 & \cellcolor{green!20}0.78 & \cellcolor{green!20}0.76 & \cellcolor{green!20}0.79 & \cellcolor{green!20}0.72 \\
& GPT 4o & \cellcolor{green!20}0.66 & \cellcolor{green!20}0.70 & \cellcolor{green!20}0.74 & \cellcolor{green!20}0.61 & \cellcolor{green!20}0.59 & \cellcolor{green!20}0.72 & \cellcolor{green!20}0.64 \\
& \cellcolor{blue!15}\textit{Average} & \cellcolor{blue!15}\textbf{0.77} & \cellcolor{blue!15}\textbf{0.81} & \cellcolor{blue!15}\textbf{0.81} & \cellcolor{blue!15}\textbf{0.75} & \cellcolor{blue!15}\textbf{0.73} & \cellcolor{blue!15}\textbf{0.80} & \cellcolor{blue!15}\textbf{0.49} \\
\midrule
\multirow{5}{*}{\rotatebox[origin=c]{90}{\textbf{VG-Cap}}} & \cellcolor{gray!15}Ground Truth & \cellcolor{gray!15}1.00 & \cellcolor{gray!15}1.00 & \cellcolor{gray!15}1.00 & \cellcolor{gray!15}1.00 & \cellcolor{gray!15}1.00 & \cellcolor{gray!15}1.00 & \cellcolor{gray!15}1.00 \\
& Claude 4.5 Sonnet & \cellcolor{green!20}0.62 & \cellcolor{green!20}0.57 & \cellcolor{green!20}0.71 & \cellcolor{green!20}0.62 & \cellcolor{green!20}0.65 & \cellcolor{green!20}0.71 & \cellcolor{green!20}0.60 \\
& Gemini 3 Pro & 0.48 & 0.47 & \cellcolor{green!20}0.55 & 0.49 & 0.43 & \cellcolor{green!20}0.53 & 0.48 \\
& GPT 4o & \cellcolor{green!20}0.52 & 0.46 & \cellcolor{green!20}0.55 & 0.47 & \cellcolor{green!20}0.53 & \cellcolor{green!20}0.54 & \cellcolor{green!20}0.55 \\
& \cellcolor{blue!15}\textit{Average} & \cellcolor{blue!15}\textbf{0.66} & \cellcolor{blue!15}\textbf{0.63} & \cellcolor{blue!15}\textbf{0.70} & \cellcolor{blue!15}\textbf{0.65} & \cellcolor{blue!15}\textbf{0.65} & \cellcolor{blue!15}\textbf{0.69} & \cellcolor{blue!15}\textbf{0.66} \\
\midrule
\multirow{5}{*}{\rotatebox[origin=c]{90}{\textbf{VG-Edit}}} & \cellcolor{gray!15}Ground Truth & \cellcolor{gray!15}-0.10 & \cellcolor{gray!15}0.10 & \cellcolor{gray!15}1.00 & \cellcolor{gray!15}1.00 & \cellcolor{gray!15}0.27 & \cellcolor{gray!15}1.00 & \cellcolor{gray!15}0.08 \\
& Claude 4.5 Sonnet & 0.29 & 0.30 & 0.49 & \cellcolor{green!20}0.53 & 0.28 & 0.45 & 0.48 \\
& Gemini 3 Pro & 0.49 & 0.47 & \cellcolor{green!20}0.54 & \cellcolor{green!20}0.57 & 0.04 & \cellcolor{green!20}0.61 & \cellcolor{green!20}0.56 \\
& GPT 4o & \cellcolor{green!20}0.59 & \cellcolor{green!20}0.61 & \cellcolor{green!20}0.61 & \cellcolor{green!20}0.69 & 0.29 & \cellcolor{green!20}0.64 & \cellcolor{green!20}0.62 \\
& \cellcolor{blue!15}\textit{Average} & \cellcolor{blue!15}\textbf{0.32} & \cellcolor{blue!15}\textbf{0.37} & \cellcolor{blue!15}\textbf{0.66} & \cellcolor{blue!15}\textbf{0.70} & \cellcolor{blue!15}\textbf{0.22} & \cellcolor{blue!15}\textbf{0.67} & \cellcolor{blue!15}\textbf{0.43} \\
\midrule
\multirow{5}{*}{\rotatebox[origin=c]{90}{\textbf{VG-Text}}} & \cellcolor{gray!15}Ground Truth & \cellcolor{gray!15}0.01 & \cellcolor{gray!15}-0.07 & \cellcolor{gray!15}-0.08 & \cellcolor{gray!15}0.19 & \cellcolor{gray!15}-0.19 & \cellcolor{gray!15}0.15 & \cellcolor{gray!15}-0.07 \\
& Claude 4.5 Sonnet & 0.16 & 0.43 & \cellcolor{green!20}0.58 & 0.21 & 0.15 & 0.23 & 0.08 \\
& Gemini 3 Pro & 0.37 & 0.42 & 0.44 & 0.48 & 0.24 & 0.37 & 0.32 \\
& GPT 4o & \cellcolor{green!20}0.50 & \cellcolor{green!20}0.71 & \cellcolor{green!20}0.63 & \cellcolor{green!20}0.58 & 0.25 & \cellcolor{green!20}0.66 & \cellcolor{green!20}0.55 \\
& \cellcolor{blue!15}\textit{Average} & \cellcolor{blue!15}\textbf{0.26} & \cellcolor{blue!15}\textbf{0.38} & \cellcolor{blue!15}\textbf{0.40} & \cellcolor{blue!15}\textbf{0.37} & \cellcolor{blue!15}\textbf{0.11} & \cellcolor{blue!15}\textbf{0.35} & \cellcolor{blue!15}\textbf{0.22} \\
\bottomrule
\end{tabular}
}
\end{table*}

\emph{VLM-as-judge} (VLMAJ) metrics have become popular because they provide strong supervision signals for subjective task assessments, especially in text and image generation tasks~\cite{manas2024improving}. \textit{Existing VLMAJ metrics do not capture the nuances of SVG code and SVG rendering.} They are also not reliable for tasks such as sketch based generation and SVG editing, where no consistent metric previously existed. For this reason we design a metric specifically tailored to the four SVG generation tasks in our benchmark.

We generate outputs from several strong baseline models and then apply carefully designed prompts to a set of powerful VLMs, both open and closed source, to obtain scores from 0 to 5 following clear evaluation criteria (see Appendix~\ref{sec:prompts-appendix}). We run the same evaluation setup with human raters and then compute Pearson correlations between VLM and human scores. This produces four task specific VLMAJ metrics, one for each task in our benchmark, providing a more faithful measure of instruction following, SVG structural correctness, and semantic alignment.

\textbf{1. Metric Development Process.} We carefully develop task-specific evaluation prompts designed to guide VLMs in assessing different aspects of SVG generation quality. For each of the four main generation tasks, we craft specialized prompts that encourage models to evaluate: (1) visual accuracy and fidelity; (2) semantic alignment with input requirements; (3) code quality and efficiency; and (4) overall aesthetic appeal.

\textbf{2. Judge Model Selection.} To identify the most reliable VLM judge, we conduct a systematic comparison across state-of-the-art models: Claude 4.5 Sonnet, Gemini 2.5 Flash, Gemini 3 Pro, Qwen2.5-VL-72B-Instruct~\citep{bai2025qwen25vl}, Qwen3-VL-235B-A22B-Instruct~\citep{bai2025qwen3vl}, and GLM 4.5 355B, covering closed-open source performance, and large-mid scale sizes.

\textbf{3. SVG Generation and VLMAJ Evaluation.} We evaluate state of the art models on the validation set (100 samples). We select Claude 4.5 Sonnet, Gemini 3 Pro, and GPT-4o, and run generation experiments on the four tasks. The resulting outputs are then scored by all VLM judges described above. We also compute scores for the ground truth SVGs, which should receive the highest ratings, providing a way to assess the overall dataset quality.

\textbf{4. Human Evaluation.} We repeat the same evaluation setup with human raters. They receive the same prompt with the specified criteria and score the generations from all models as well as the ground truth data. A total of 17 human evaluators participated, all technical engineers or AI and design experts, producing around 674 ratings used to correlate each VLM with human judgment.

\textbf{5. Correlation Validation and VLMAJ Selection.} We compute Pearson correlation coefficients between human judgments and each candidate VLM judge for every task and report the results in Table~\ref{tab:vlm_correlation}. We also include average validation scores for the three generation models in Table~\ref{tab:averages}, showing both human ratings and VLM evaluations.
Ground Truth acts as a reliable anchor only for VG Sketch and VG Cap, where human agreement is high due to clearer visual semantics. For VG Edit and VG Text, correlations drop even on perfect examples, indicating that these tasks contain more structural ambiguity and are inherently harder to evaluate with full consensus. This further motivates the need for robust automatic judges tailored to each task.
The correlation results highlight clear preferences among VLM judges. Gemini Flash and Gemini 3 Pro provide the strongest alignment with human ratings in VG Sketch, and Gemini 3 Pro also achieves the highest correlation in VG Cap. For VG Edit, which is the most challenging task, Gemini 3 Pro and GPT 5.1 stand out as the only reliable options, with GPT 5.1 showing a slight advantage. For VG Text, Gemini Flash ranks highest, with GPT 5.1 again performing consistently.
Qwen3 VL 235B emerges as the most stable open source option, performing well across VG Sketch, VG Cap, and VG Edit, with the main weakness appearing in VG Text.
Based on these findings, we select Gemini 3 Pro as the primary VLMAJ judge for VG Sketch, VG Cap, and VG Text. For VG Edit, we use GPT 5.1, which shows the strongest alignment with human judgments on this task.

\section{Additional Qualitative Results}

\begin{figure*}[t]
  \centering
  \includegraphics[width=0.8\textwidth]{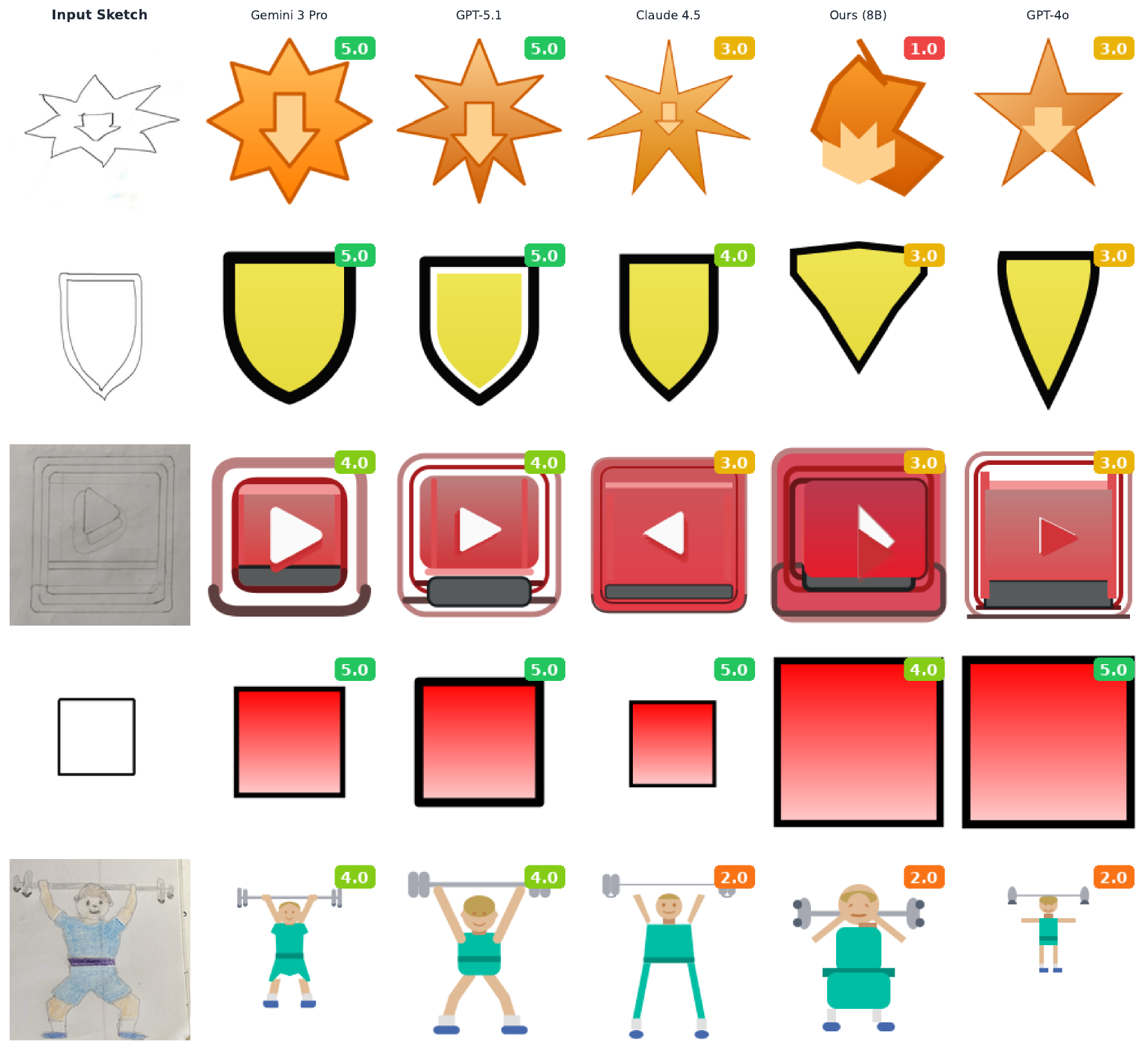}
  \caption{\textbf{VG-Sketch Qualitative Results.} The leftmost column displays the input raster sketch, followed by the outputs from top-performing models. Gemini 3 Pro demonstrates superior fidelity in preserving topological structure compared to GPT-5.1 and others.}
  \label{fig:vg-sketch-qualitative}
\end{figure*}

\begin{figure*}[t]
  \centering
  \includegraphics[width=0.8\textwidth]{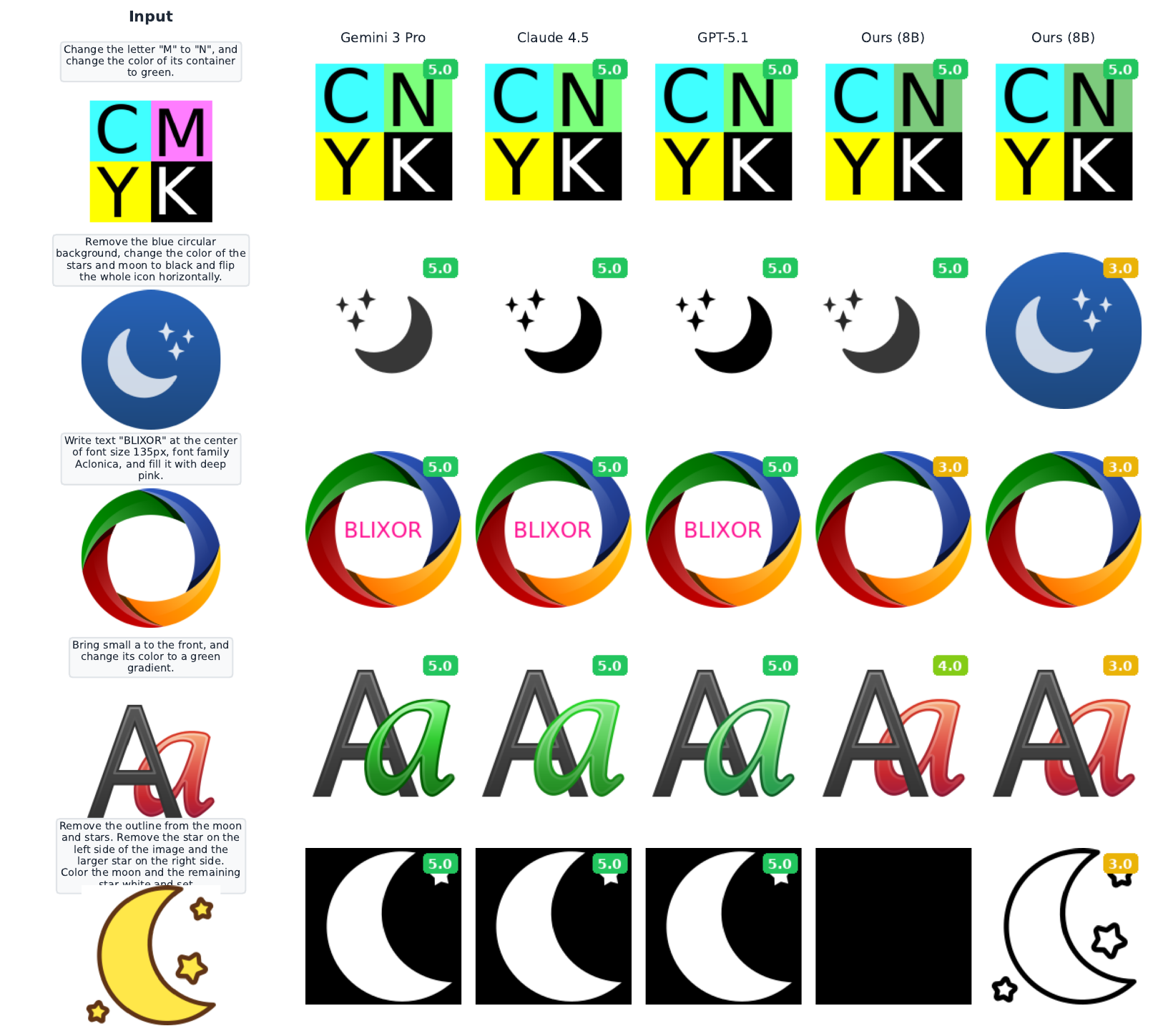}
  \caption{\textbf{VG-Edit Qualitative Results.} Left to right: natural language edit instruction, input SVG, and model outputs. Gemini 3 Pro, Claude 4.5 Sonnet, and GPT5.1 effectively execute complex semantic modifications, whereas our trained models struggle to follow some multi-step edits.}
  \label{fig:vg-edit-qualitative}
\end{figure*}

We provide additional qualitative figures, including the VG-Sketch and VG-Edit result visualizations referenced in the main text (\Cref{fig:vg-sketch-qualitative,fig:vg-edit-qualitative}).

\begin{figure*}
\centering
\includegraphics[width=0.8\textwidth, trim=0 600 0 0,clip]{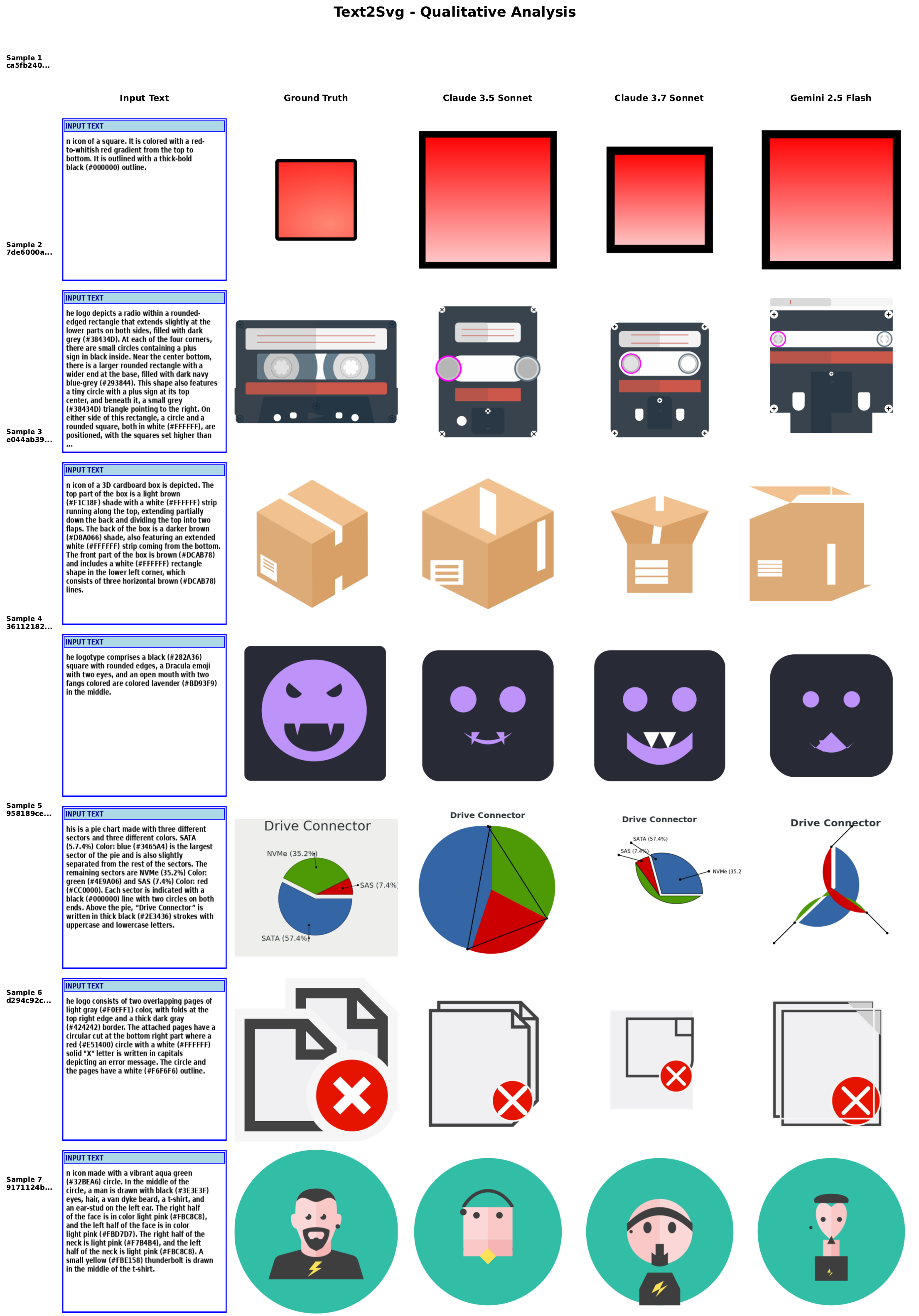}
\caption{Qualitative analysis of Text2SVG generation results. The figure shows examples of text2SVG generation across different model performances. Examples demonstrate successful generations with accurate semantic understanding and geometric representation, as well as common failure modes including incorrect primitive usage, semantic misunderstanding, and incomplete shape representations.}
\label{fig:text2svg_qualitative}
\end{figure*}

\begin{figure*}
\centering
\includegraphics[width=0.95\textwidth,trim=0 600 0 0,clip]{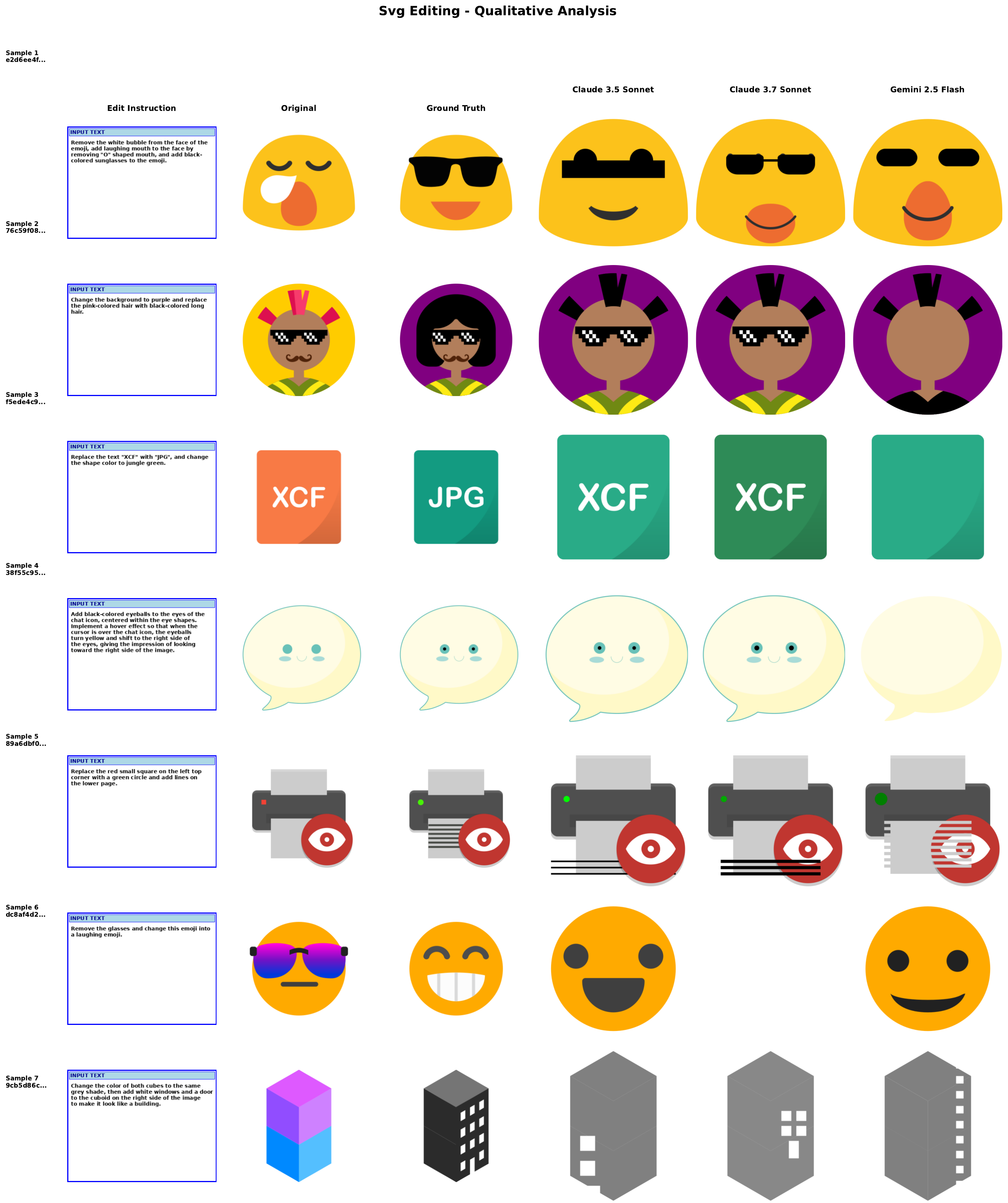}
\caption{\textbf{Qualitative analysis of SVG editing with natural language instructions}. The figure demonstrates model performance on various editing tasks including color changes, geometric transformations, and structural modifications. Examples show input SVG (left), editing instruction (center), and generated output (right). Successful cases highlight accurate instruction parsing and precise SVG manipulation, while failure cases reveal challenges in understanding complex instructions and maintaining visual coherence.}
\label{fig:svg_editing_qualitative}
\end{figure*}

\textbf{Failure Analysis.} Figures~\ref{fig:vg-edit-qualitative}, \ref{fig:text2svg_qualitative}, and \ref{fig:svg_editing_qualitative} qualitatively illustrate incorrect primitives, semantic misunderstanding, incomplete shapes or edits, and loss of coherence rather than population-level frequencies.

\begin{figure*}
  \centering 
  \includegraphics[width=0.8\linewidth,trim={0 140pt 0 0},clip]{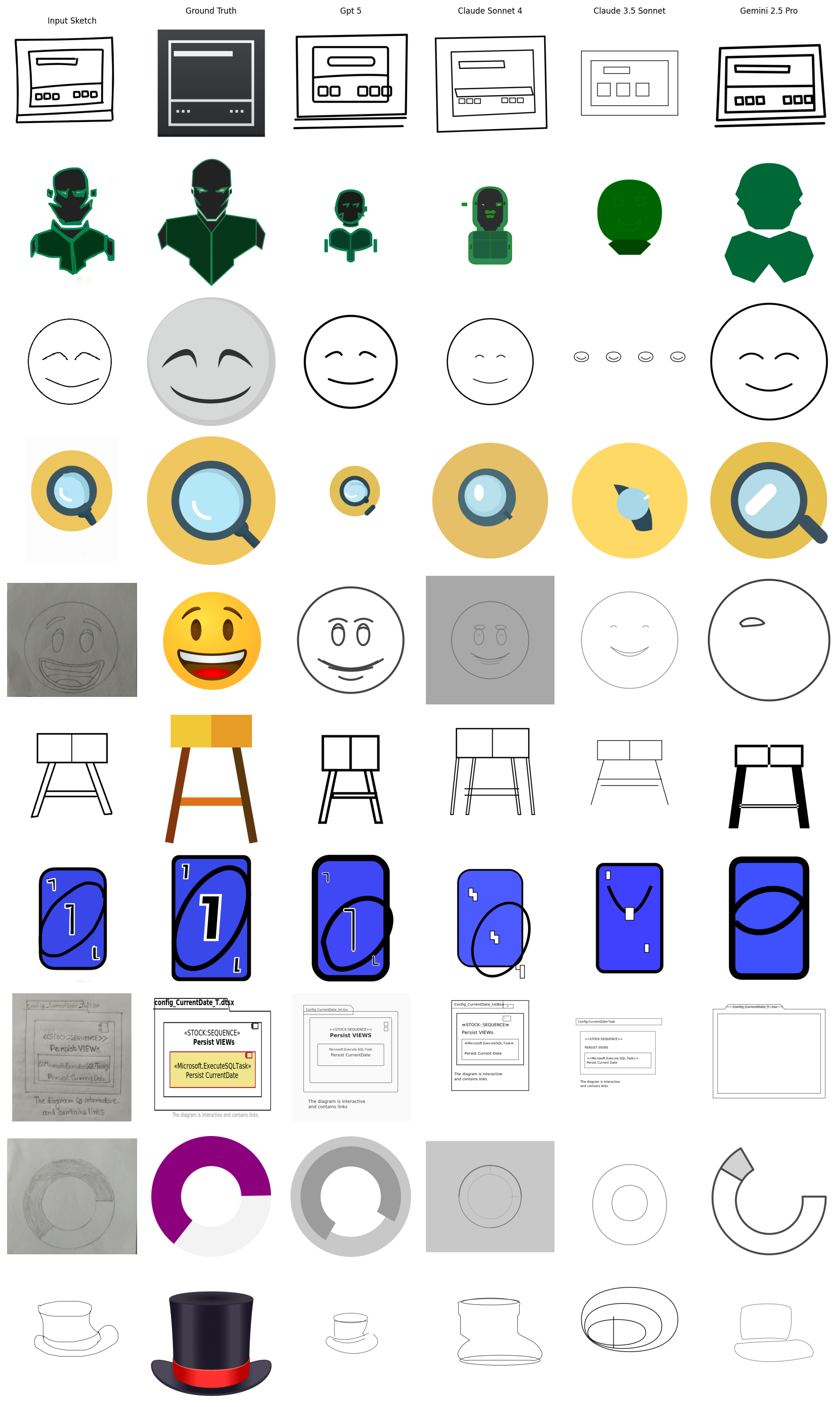} 
  \caption{Visualization of test performance on the Sketch2SVG task. When the input sketch lacks color, models tend not to introduce new colors. In contrast, when color is present in the sketch, models successfully reproduce it in the generated SVG.}
  \label{fig:sketch-app} 
  \end{figure*}

\begin{figure*}
\centering
\includegraphics[width=0.95\textwidth,trim=0 500 0 0,clip]{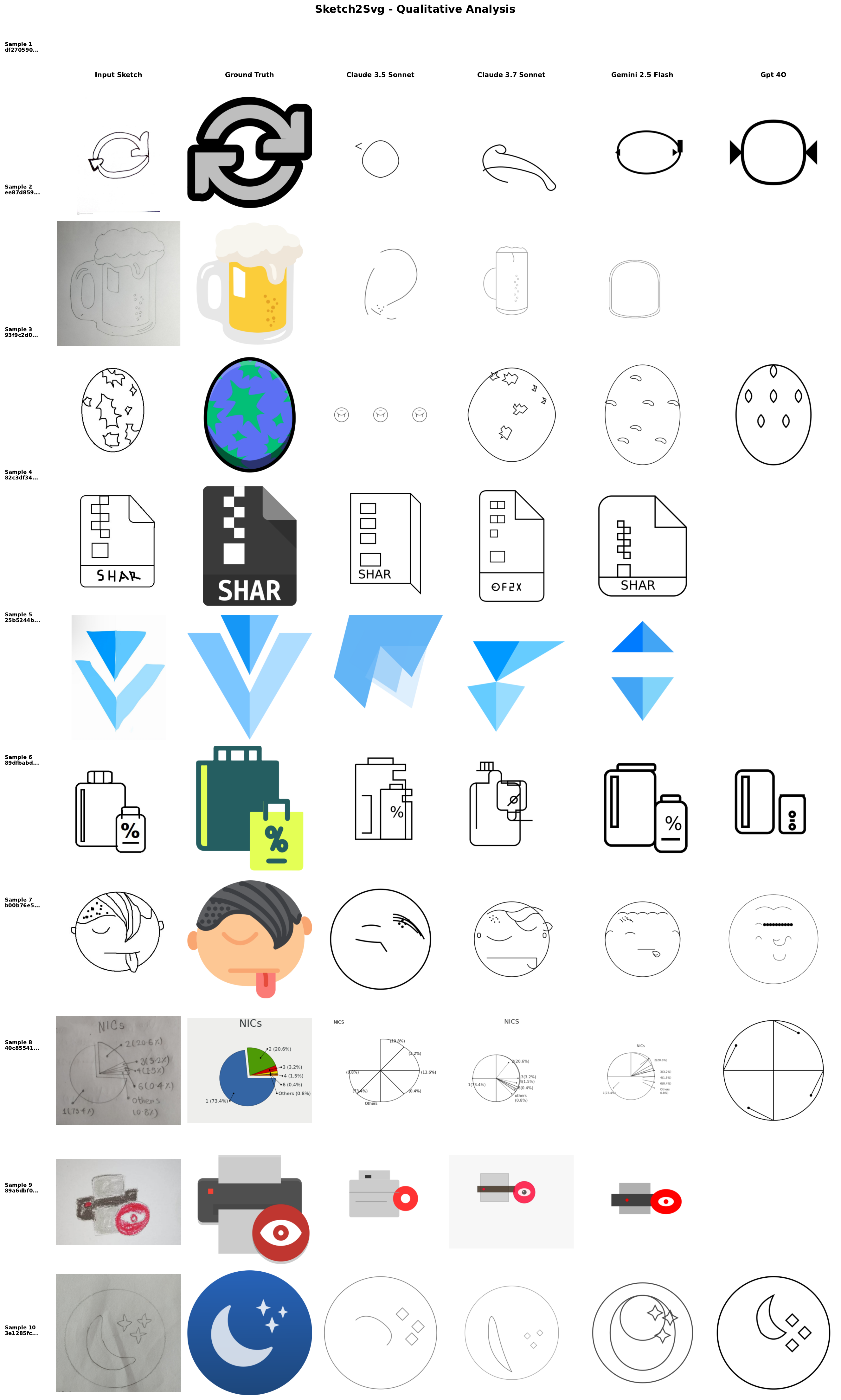}
\caption{\textbf{Qualitative analysis of Sketch2SVG generation results}. The figure illustrates model performance in converting hand-drawn sketches to clean SVG code. Examples display input sketches (left), ground truth SVG (second column), and model-generated SVGs (rest of the columns).}
\label{fig:sketch2svg_qualitative}
\end{figure*}

\begin{figure*}
\centering
\includegraphics[width=0.95\textwidth,trim=0 0 0 0,clip]{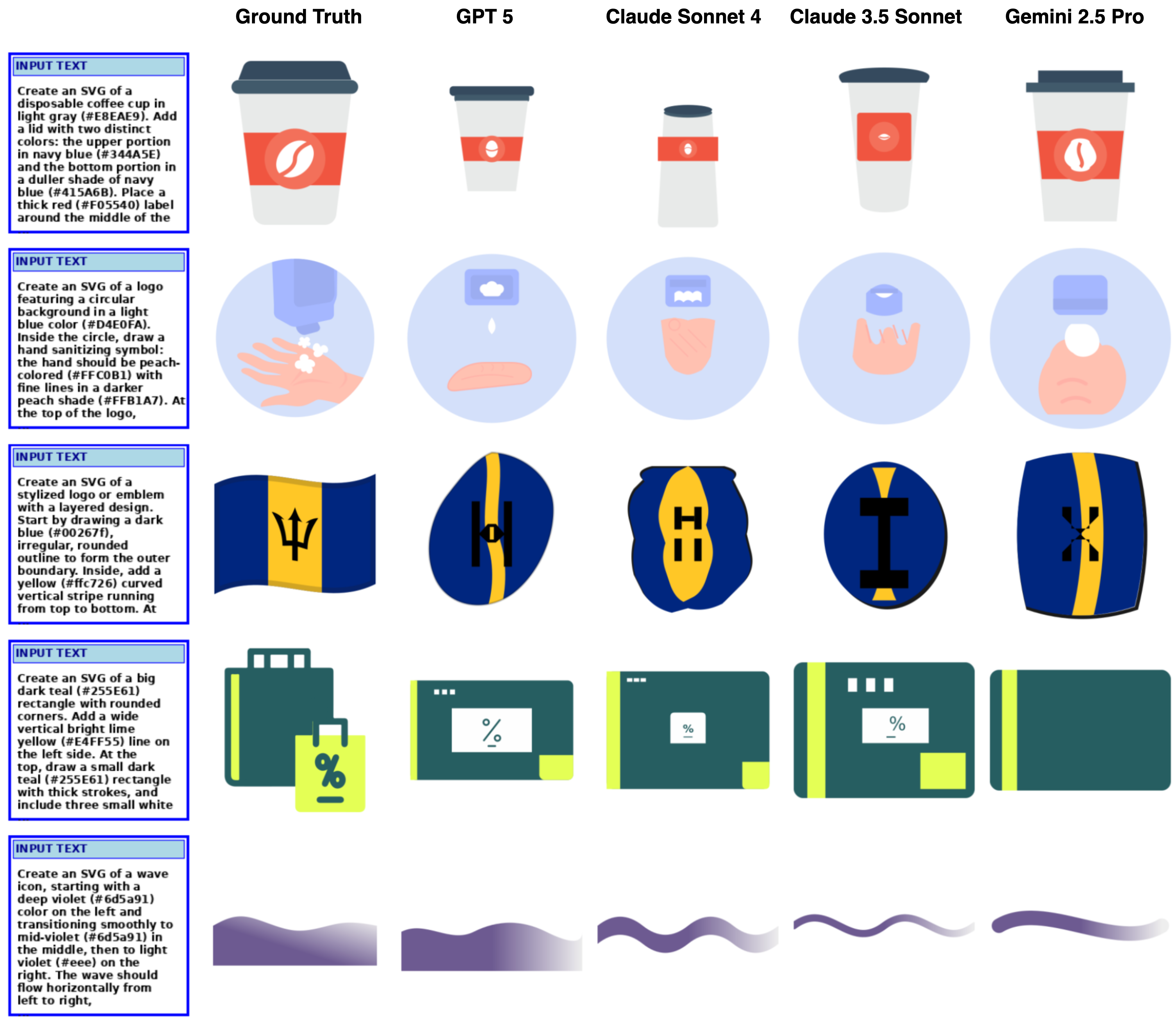}
\caption{\textbf{Qualitative analysis of Text2SVG generation results}.}
\label{fig:text2svg}
\end{figure*}

\begin{table*}[t]
  \centering
  \caption{Comparison of SVG datasets and benchmarks. \textbf{VectorGym (Ours)} is the only benchmark combining multi-task evaluation with human-verified quality. Note: Size is reported in number of SVG samples.}
  \label{tab:svg_datasets_extended}
  \resizebox{\textwidth}{!}{%
  \begin{tabular}{@{}l c r l l l@{}}
  \toprule
  \textbf{Dataset} & \textbf{Year} & \textbf{Size} & \textbf{Content Types} & \textbf{Tasks} & \textbf{Annotation} \\
  \midrule
  \textbf{VG-Sketch (Ours)} & 2025 & 6.5k & Icons, Fonts, Diagrams, Emojis & Sketch-to-SVG & Human \\
  \textbf{VG-Text2SVG (Ours)} & 2025 & 6.5k & Icons, Diagrams, Emojis, Fonts & Text-to-SVG & Human \\
  \textbf{VG-Edit (Ours)} & 2025 & 6.5k & Diverse & SVG Editing & Human \\
  \midrule
  SVG-Stack & 2025 & 2.3M & Diverse (Icons, Logos, Diagrams) & SVG Corpus & Unlabeled \\
  Text2SVG-Stack & 2025 & 2.2M & Diverse (Paired Texts and SVGs) & Text-to-SVG & Synthetic Captions \\
  SVG-Fonts & 2025 & 1.9M & Fonts, Glyphs & SVG Corpus & Unlabeled \\
  SVG-Icons & 2025 & 89k & Icons & SVG Corpus & Unlabeled \\
  SVG-Emoji & 2025 & 10k & Emojis & SVG Corpus & Unlabeled \\
  \midrule
  MMSVG-2M & 2025 & 2.0M & Icons, Illustrations, Characters & Image/Text-to-SVG & Mixed (Web + Syn.) \\
  UniSVG & 2025 & 525k & Unified Multi-domain & Gen. \& Understanding & Mixed \\
  SVGX-SFT-1M & 2025 & 1.0M & Diverse (Instr.$\leftrightarrow$SVG) & Instruction Following & Synthetic (LLM) \\
  SVG-1M (SVGen) & 2025 & 1.0M & Icons & Image/Text-to-SVG & Synthetic (LLM) \\
  FIGR-SVG & 2025 & 1.3M & Icons & Text/Image-to-SVG & Converted + Syn. \\
  DeepSVG Dataset & 2020 & 100k & Icons & SVG Generation & Curated \\
  \midrule
  SVGenius & 2025 & 2.4k & Diverse & Understanding \& Editing & Human-verified \\
  VGBench & 2024 & 10k & Multi-format (SVG, TikZ, Graphviz) & Understanding \& Gen. & Synthetic + Verified \\
  SVGEditBench v2 & 2025 & 1.7k & Emojis, Icons & SVG Editing & Synthetic Prompts \\
  VectorEdits & 2025 & 270k & Diverse & SVG Editing (Guided) & Synthetic (VLM) \\
  \midrule
  Quick Draw! & 2017 & 50M & Sketches & Sketch Recognition & Human \\
  IconDesc & 2024 & 1.4k & UI Icons & Captioning (Alt-text) & Human \\
  \bottomrule
  \end{tabular}%
  }
\end{table*}

\section{Prompts}
\label{sec:prompts-appendix}

In this section we present all the prompts used throughout the paper. We designed task specific prompts for SVG generation across the four main tasks, and we also crafted evaluation prompts that guide models to score outputs in a way that captures the semantic quality of the SVG rather than focusing on pixel based visual features. We validated the effectiveness of these evaluation prompts through a correlation analysis, shown in table~\ref{tab:vlm_correlation}.

\begin{table*}[t]
\centering
\small
\caption{\textbf{Scores for human evaluation and VLMAJ.} We show average scores by generator model and VLM judge across different tasks.}
\label{tab:averages}
\resizebox{\textwidth}{!}{
\begin{tabular}{@{}llcccccccc@{}}
\toprule
\textbf{Task} & \textbf{Generator} & \textbf{Human} & \multicolumn{7}{c}{\textbf{Models used as Judges}} \\
\cmidrule(lr){4-10}
 & & & Claude 4.5 Sonnet & Gemini 2.5 Flash & Gemini 3 Pro & GPT 5.1 & Qwen2.5VL 72B & Qwen3.VL 235B & GLM4.5 355B \\
\midrule
\multirow{4}{*}{\rotatebox[origin=c]{90}{\textbf{VG-Sketch}}} & GPT 4o & 2.57 & 2.79 & 2.46 & 2.43 & 3.16 & 3.10 & 2.79 & 2.05 \\
& Claude 4.5 Sonnet & 2.88 & 3.22 & 2.91 & 2.81 & 3.57 & 3.70 & 3.34 & 2.46 \\
& Gemini 3 Pro & \textbf{3.63} & \textbf{3.55} & \textbf{3.41} & \textbf{3.49} & \textbf{3.72} & \textbf{3.91} & \textbf{3.74} & \textbf{2.75} \\
& \cellcolor{gray!15}Ground Truth & \cellcolor{gray!15}4.79 & \cellcolor{gray!15}5.00 & \cellcolor{gray!15}5.00 & \cellcolor{gray!15}5.00 & \cellcolor{gray!15}5.00 & \cellcolor{gray!15}5.00 & \cellcolor{gray!15}5.00 & \cellcolor{gray!15}4.97 \\
\midrule
\multirow{4}{*}{\rotatebox[origin=c]{90}{\textbf{VG-Cap}}} & GPT 4o & 2.90 & 2.15 & 0.84 & 2.26 & 2.21 & 1.27 & 1.74 & 1.20 \\
& Claude 4.5 Sonnet & 3.67 & 2.60 & 1.43 & 2.86 & 2.87 & 1.80 & 2.19 & 1.87 \\
& Gemini 3 Pro & \textbf{3.95} & \textbf{2.73} & \textbf{1.69} & \textbf{3.20} & \textbf{3.12} & \textbf{1.81} & \textbf{2.35} & \textbf{2.03} \\
& \cellcolor{gray!15}Ground Truth & \cellcolor{gray!15}4.67 & \cellcolor{gray!15}5.00 & \cellcolor{gray!15}5.00 & \cellcolor{gray!15}5.00 & \cellcolor{gray!15}5.00 & \cellcolor{gray!15}5.00 & \cellcolor{gray!15}5.00 & \cellcolor{gray!15}5.00 \\
\midrule
\multirow{4}{*}{\rotatebox[origin=c]{90}{\textbf{VG-Edit}}} & GPT 4o & 2.22 & 2.17 & 2.19 & 2.62 & 2.78 & 2.32 & 3.01 & 2.30 \\
& Claude 4.5 Sonnet & 3.35 & 3.15 & 3.23 & 3.45 & 3.79 & 2.89 & 3.88 & 3.16 \\
& Gemini 3 Pro & \textbf{4.07} & \textbf{3.46} & \textbf{3.54} & \textbf{3.78} & \textbf{4.11} & \textbf{3.16} & \textbf{4.12} & \textbf{3.45} \\
& \cellcolor{gray!15}Ground Truth & \cellcolor{gray!15}4.41 & \cellcolor{gray!15}4.18 & \cellcolor{gray!15}4.46 & \cellcolor{gray!15}5.00 & \cellcolor{gray!15}5.00 & \cellcolor{gray!15}4.18 & \cellcolor{gray!15}5.00 & \cellcolor{gray!15}4.70 \\
\midrule
\multirow{4}{*}{\rotatebox[origin=c]{90}{\textbf{VG-Text}}} & GPT 4o & 2.19 & 3.23 & 2.69 & 3.40 & 3.52 & 2.72 & 3.14 & 3.28 \\
& Claude 4.5 Sonnet & 2.73 & \textbf{4.11} & 3.52 & 4.36 & \textbf{4.33} & 3.20 & 3.90 & \textbf{4.22} \\
& Gemini 3 Pro & \textbf{3.33} & 4.10 & \textbf{3.58} & \textbf{4.55} & 4.24 & \textbf{3.27} & \textbf{4.04} & 4.17 \\
& \cellcolor{gray!15}Ground Truth & \cellcolor{gray!15}4.66 & \cellcolor{gray!15}4.18 & \cellcolor{gray!15}3.78 & \cellcolor{gray!15}4.87 & \cellcolor{gray!15}4.56 & \cellcolor{gray!15}3.49 & \cellcolor{gray!15}4.24 & \cellcolor{gray!15}4.23 \\
\bottomrule
\end{tabular}
}
\end{table*}


\begin{figure}
    \centering
    \includegraphics[width=\linewidth]{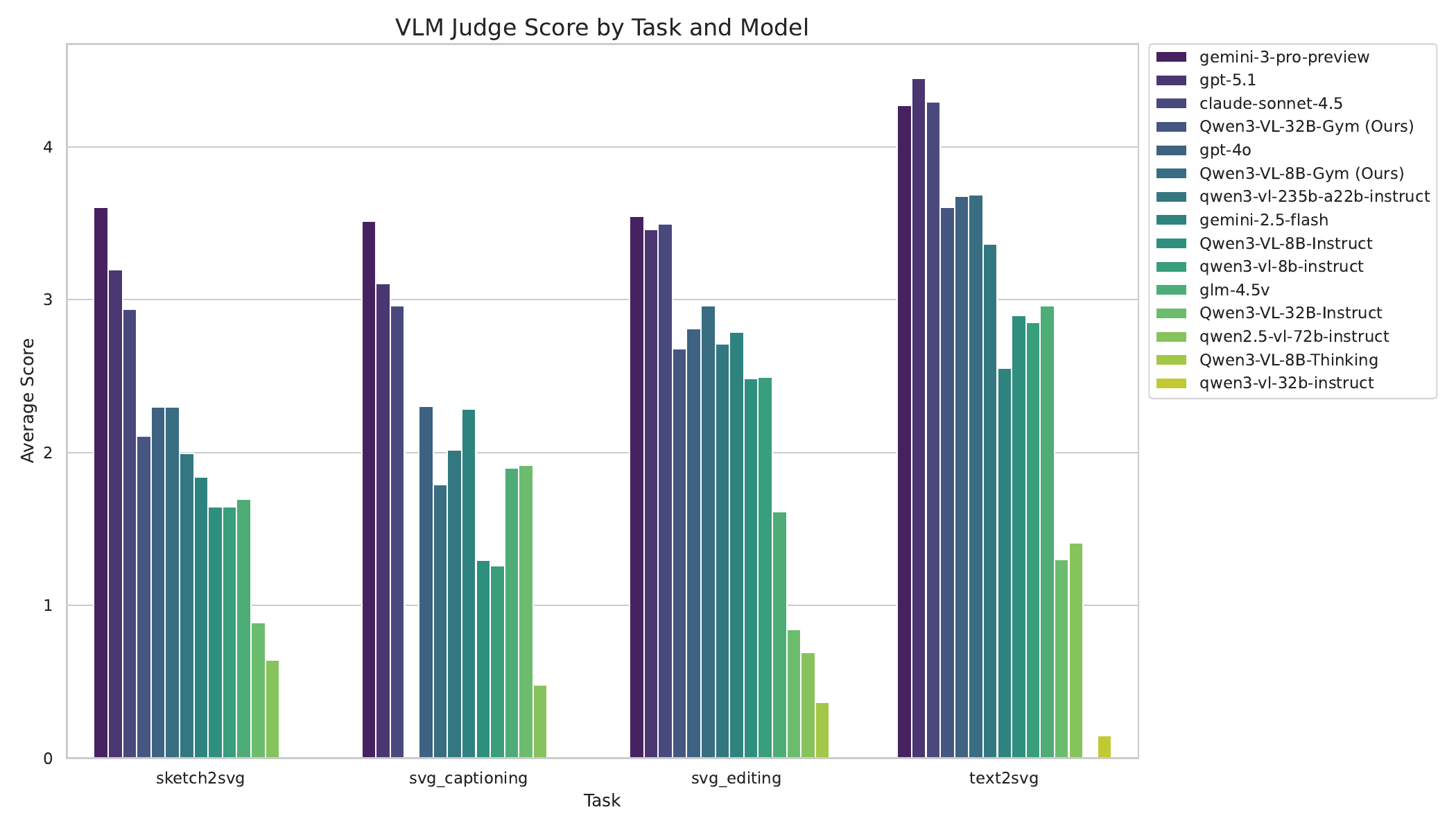}
    \caption{VLM-as-judge scores for different tasks and models.}
    \label{fig:judge-scores}
\end{figure}

\subsection{VLM-as-a-Judge Prompts}

\begin{PromptBox}[colback=text2svgBack, colframe=text2svgAccent, colbacktitle=text2svgAccent, borderline west={2pt}{0pt}{text2svgAccent}]{VLM-as-a-Judge (Text2SVG)}
\label{prompt:llm-judge-txt2svg}
You are a concise evaluator of text-to-SVG faithfulness. 
Judge how well a generated SVG image matches its textual description. 
Focus primarily on semantic content (what is shown), not exact wording or artistic style. 
Do not use world knowledge; base your judgment only on what the text states and what is visible.

\textbf{Evaluation Instructions:} Compare the generated image to the TEXT description. 
Judge semantic/visual meaning, not exact wording.

\textbf{Rules:}
\begin{itemize}
    \item Focus on the presence and configuration of the main objects, their attributes (shape, rough size, main color), spatial relations, and overall layout.
    \item Accept paraphrases and synonyms; do not require exact wording.
    \item Numbers, counts, colors, attributes, and relations are important: penalize clear mismatches, but do not over-penalize small deviations when the overall scene clearly matches the text.
    \item Penalize unsupported or clearly contradictory visual details (hallucinations) more than omissions.
    \item Consider image quality, clarity, and coherence as a secondary factor: when semantic match is similar, prefer clearer and better-formed SVGs.
    \item Ignore minor stylistic differences (line style, thickness, minor artifacts), casing, and punctuation.
    \item Do not use world knowledge; compare only what the TEXT states and what is visible.
\end{itemize}

\textbf{Text Description:} \texttt{\{caption\}}

\textbf{Scoring Rubric (0--5):}
\begin{itemize}
    \item \textbf{5:} Very strong match; main objects, layout, and key attributes align with the text; only small local details differ; no strong contradictions.
    \item \textbf{4:} Good match; overall scene corresponds to the text with only minor issues.
    \item \textbf{3:} Partial match; several core elements align, but some important detail is missing, wrong, or extra.
    \item \textbf{2:} Weak match; topic is similar but multiple important errors, omissions, or hallucinated details.
    \item \textbf{1:} Minimal overlap; only a very generic aspect matches.
    \item \textbf{0:} Unrelated or contradicts core facts.
\end{itemize}

Output ONLY the integer score (0--5). No words, no JSON, no explanations.

\textbf{Output format:}
\begin{verbatim}
<0-5>

\end{verbatim}

\end{PromptBox} 

\begin{PromptBox}[colback=sketch2svgBack, colframe=sketch2svgAccent, colbacktitle=sketch2svgAccent, borderline west={2pt}{0pt}{sketch2svgAccent}]{VLM-as-a-Judge (Sketch2SVG)}
\label{prompt:llm-judge-sketch2svg}
You are a concise evaluator of sketch-to-image similarity. 
Judge how well the generated image preserves the semantic content and structure of the input sketch.

\textbf{Evaluation Instructions:} Compare the PREDICTION image directly to the GROUND-TRUTH image. 
Judge semantic similarity and preservation of visual content, not artistic style.

\textbf{Rules:}
\begin{itemize}
    \item Focus on the main objects, their presence or absence, shapes, sizes, colors, and spatial relations.
    \item Treat numbers, counts, colors, attributes, and relative positions as important; penalize clear mismatches.
    \item Penalize added elements that are not present in the ground-truth image (hallucinations) more than small omissions.
    \item Penalize missing or significantly altered key elements more than minor stylistic or rendering differences.
    \item Ignore small artifacts, minor shading/texture differences, or slight geometric deviations if the overall content clearly matches.
    \item Do not use world knowledge; compare only what is visible in the GROUND-TRUTH and PREDICTION images.
\end{itemize}

\textbf{Inputs:}
\begin{itemize}
    \item \textbf{GROUND-TRUTH image:} the target image.
    \item \textbf{PREDICTION image:} the model-generated image to be evaluated.
\end{itemize}

\textbf{Scoring Rubric (0--5):}
\begin{itemize}
    \item \textbf{5:} Very strong match; all main objects and key attributes align; only small local or stylistic differences.
    \item \textbf{4:} Good match; overall scene clearly corresponds, with one or a few noticeable but non-critical differences.
    \item \textbf{3:} Partial match; several core elements align, but some important details are missing, wrong, or extra.
    \item \textbf{2:} Weak match; topic is similar, but multiple important elements are missing, incorrect, or hallucinated.
    \item \textbf{1:} Minimal overlap; only very generic aspects (e.g., rough layout or general type of scene) match.
    \item \textbf{0:} Unrelated or clearly contradicts the ground-truth (wrong main objects, layout, or overall scene).
\end{itemize}

Output ONLY the integer score (0--5). No words, no JSON, no explanations.

\textbf{Output format:}
\begin{verbatim}
<0-5>

\end{verbatim}
\end{PromptBox}

\begin{PromptBox}[colback=svgeditBack, colframe=svgeditAccent, colbacktitle=svgeditAccent, borderline west={2pt}{0pt}{svgeditAccent}]{VLM-as-a-Judge (SVG-Editing)}
\label{prompt:llm-judge-svg-edit}
You are a concise evaluator for image editing results. 
Judge how well a PREDICTION image matches a GROUND-TRUTH image. 
Do not use world knowledge; rely only on the visible content of the two images.

\textbf{Evaluation Instructions:} Compare the PREDICTION image directly to the GROUND-TRUTH image. 
Judge semantic similarity and preservation of visual content, not artistic style.

\textbf{Rules:}
\begin{itemize}
    \item Focus on the main objects, their presence or absence, shapes, sizes, colors, and spatial relations.
    \item Treat numbers, counts, colors, attributes, and relative positions as important; penalize clear mismatches.
    \item Penalize added elements that are not present in the ground-truth image (hallucinations) more than small omissions.
    \item Penalize missing or significantly altered key elements more than minor stylistic or rendering differences.
    \item Ignore small artifacts, minor shading/texture differences, or slight geometric deviations if the overall content clearly matches.
    \item Do not use world knowledge; compare only what is visible in the GROUND-TRUTH and PREDICTION images.
\end{itemize}

\textbf{Inputs:}
\begin{itemize}
    \item \textbf{GROUND-TRUTH image:} the target image.
    \item \textbf{PREDICTION image:} the model-generated image to be evaluated.
\end{itemize}

\textbf{Scoring Rubric (0--5):}
\begin{itemize}
    \item \textbf{5:} Very strong match; all main objects and key attributes align; only small local or stylistic differences.
    \item \textbf{4:} Good match; overall scene clearly corresponds, with one or a few noticeable but non-critical differences.
    \item \textbf{3:} Partial match; several core elements align, but some important details are missing, wrong, or extra.
    \item \textbf{2:} Weak match; topic is similar, but multiple important elements are missing, incorrect, or hallucinated.
    \item \textbf{1:} Minimal overlap; only very generic aspects (e.g., rough layout or general type of scene) match.
    \item \textbf{0:} Unrelated or clearly contradicts the ground-truth (wrong main objects, layout, or overall scene).
\end{itemize}

Output ONLY the integer score (0--5). No words, no JSON, no explanations.

\textbf{Output format:}
\begin{verbatim}
<0-5>

\end{verbatim}

\end{PromptBox}

\begin{PromptBox}[colback=svgcaptionBack, colframe=svgcaptionAccent, colbacktitle=svgcaptionAccent, borderline west={2pt}{0pt}{svgcaptionAccent}]{VLM-as-a-Judge (SVG-Captioning)}
\label{prompt:llm-judge-svg-caption}
You are a concise evaluator of caption similarity. 
Compare a PREDICTION caption to a GROUND-TRUTH caption (no image). 
Judge semantic meaning, not exact wording.

\textbf{Rules:}
\begin{itemize}
    \item Accept paraphrases and synonyms.
    \item Treat numbers, counts, colors, attributes, relations, and negation as strict.
    \item Penalize unsupported or contradictory details (hallucinations) more than omissions.
    \item Ignore casing and punctuation (except negation words like ``no/not/without'').
    \item Do not use world knowledge; compare only what the texts state.
\end{itemize}

\textbf{Scoring (return a single integer 0--5):}
\begin{itemize}
    \item \textbf{5}: Semantically equivalent or near-paraphrase; all key facts align; no contradictions.
    \item \textbf{4}: Very close; only a minor detail missing/different; no contradictions.
    \item \textbf{3}: Partially correct; several core elements match but some important detail is missing.
    \item \textbf{2}: Weak overlap; multiple important errors or added unsupported specifics.
    \item \textbf{1}: Minimal overlap; only a very generic element matches.
    \item \textbf{0}: Unrelated or contradicts core facts (e.g., negation flip, wrong main objects/actions).
\end{itemize}

Output ONLY the integer score (0--5). No words, no JSON, no explanations.

\textbf{Output format:}
\begin{verbatim}
<0-5>

\end{verbatim}
\end{PromptBox}

\subsection{SVG Generation Prompts}

\begin{PromptBox}[colback=text2svgBack, colframe=text2svgAccent, colbacktitle=text2svgAccent, borderline west={2pt}{0pt}{text2svgAccent}]{Text2SVG Generation}
\label{prompt:generation-text2svg}
You are an expert in generating SVG representations of textual descriptions. 

Follow these steps carefully:
\begin{enumerate}
    \item Analyze the given text input and identify the key visual elements it describes.
    \item Convert the description into a minimal and clear SVG representation using basic SVG shapes such as \texttt{<rect>}, \texttt{<circle>}, \texttt{<line>}, and \texttt{<path>}.
    \item Ensure the SVG design is simple, scalable, and directly represents the input text.
    \item Do not include any additional text, explanations, comments, or formatting---only output valid SVG code.
    \item The output must be a complete SVG document, starting with \texttt{<svg>} and ending with \texttt{</svg>}.
\end{enumerate}

\textbf{*** textual descriptions***}

-- textual descriptions

\textbf{*** REASONING***}

Let's think step by step then output the svg.
First, wrap your detailed reasoning process in \texttt{<think>} and \texttt{</think>} tags. In your reasoning, describe your approach in natural language WITHOUT showing code examples.
Then, output the complete SVG code directly after the closing \texttt{</think>} tag (NO markdown wrapper, NO \texttt{```xml} or \texttt{```svg} tags).
Your reasoning should consider: concept sketching, canvas planning, shape decomposition, coordinate calculation, styling and color, symbolism or metaphor, and final assembly.

IMPORTANT: After \texttt{</think>}, output ONLY the raw SVG starting with \texttt{<svg} and ending with \texttt{</svg>}. Do NOT use markdown code blocks or wrap in \texttt{```xml} or \texttt{```svg}.

\end{PromptBox}

\begin{PromptBox}[colback=sketch2svgBack, colframe=sketch2svgAccent, colbacktitle=sketch2svgAccent, borderline west={2pt}{0pt}{sketch2svgAccent}]{Sketch2SVG Generation}
\label{prompt:generation-sketch2svg}
You are an expert in generating SVG from a hand-drawn sketch plus a brief description.

\textbf{*** GOALS ***}
\begin{itemize}
    \item \textbf{Semantic match}: faithfully reflect the sketch, using the description to clarify ambiguous parts; include all and only the intended elements, attributes, and relationships.
    \item \textbf{Validity + code quality}: produce a parsable SVG with concise primitives and a tidy, readable structure.
    \item \textbf{Visual fidelity}: preserve essential contours, proportions, and layout; if gradients, shadows, or outlines are mentioned, implement them minimally.
\end{itemize}

\textbf{*** PROCEDURE ***}
\begin{enumerate}
    \item Examine the sketch to identify primary shapes, contours, and alignment; use the description to resolve labels, counts, and styling cues.  
    \item Decompose the scene into basic SVG shapes (\texttt{<rect>}, \texttt{<circle>}, \texttt{<ellipse>}, \texttt{<line>}, \texttt{<polygon>}, \texttt{<polyline>}, \texttt{<path>}), simplifying strokes and curves where appropriate.  
    \item Translate relative placements and sizes from the sketch into a coherent coordinate system and consistent stroke/fill attributes.  
    \item Apply only the necessary styling (strokes, fills, minimal effects) specified or implied by the sketch and description.  
    \item Output only valid SVG code as a complete document enclosed by \texttt{<svg>} and \texttt{</svg>}.
\end{enumerate}

\textbf{*** SVG Description ***}

-- svg description

\textbf{*** REASONING***}

Let's think step by step then output the svg.
First, wrap your detailed reasoning process in \texttt{<think>} and \texttt{</think>} tags. In your reasoning, describe your approach in natural language WITHOUT showing code examples.
Then, output the complete SVG code directly after the closing \texttt{</think>} tag (NO markdown wrapper, NO \texttt{```xml} or \texttt{```svg} tags).
Your reasoning should consider: concept sketching, canvas planning, shape decomposition, coordinate calculation, styling and color, symbolism or metaphor, and final assembly.

IMPORTANT: After \texttt{</think>}, output ONLY the raw SVG starting with \texttt{<svg} and ending with \texttt{</svg>}. Do NOT use markdown code blocks or wrap in \texttt{```xml} or \texttt{```svg}.

\end{PromptBox}

\begin{PromptBox}[colback=svgeditBack, colframe=svgeditAccent, colbacktitle=svgeditAccent, borderline west={2pt}{0pt}{svgeditAccent}]{SVG Editing Generation}
\label{prompt:generation-svg-editing}
You are an expert in editing SVG images based on text instructions. 

Follow these steps carefully:
\begin{enumerate}
    \item Analyze the original SVG and the editing instruction.
    \item Apply the requested modifications while preserving the overall structure.
    \item Ensure the edited SVG is valid and well-formed.
    \item Do not include any additional text, explanations, comments, or formatting---only output valid SVG code.
    \item The output must be a complete SVG document, starting with \texttt{<svg>} and ending with \texttt{</svg>}.
\end{enumerate}

\textbf{Original SVG:}

-- svg code

\textbf{Editing Instruction:}
\begin{quote}
Reduce the image size and add a kite string extending from the bottom-right corner to make it look like a kite.
\end{quote}

\textbf{*** REASONING***}

Let's think step by step then output the edited svg.  
First, wrap your detailed reasoning process in \texttt{<think>} and \texttt{</think>} tags. In your reasoning, describe your approach in natural language WITHOUT showing code examples.  
Then, output the complete SVG code directly after the closing \texttt{</think>} tag (NO markdown wrapper, NO \texttt{```xml} or \texttt{```svg} tags).  
Your reasoning should consider: parsing the instruction, identifying target elements, determining minimal required changes, preserving unmodified elements, and validating the result.

IMPORTANT: After \texttt{</think>}, output ONLY the raw SVG starting with \texttt{<svg} and ending with \texttt{</svg>}. Do NOT use markdown code blocks or wrap in \texttt{```xml} or \texttt{```svg}.

\end{PromptBox}

\begin{PromptBox}[colback=svgcaptionBack, colframe=svgcaptionAccent, colbacktitle=svgcaptionAccent, borderline west={2pt}{0pt}{svgcaptionAccent}]{SVG Captioning Generation}
\label{prompt:generation-svg-caption}

You are an expert at describing SVG images. 
Given an SVG, provide a clear and concise caption that describes the visual elements, their colors, positions, and any notable features. 
Focus on what someone would see when looking at the rendered SVG.

\textbf{SVG:} \texttt{\{svg\}}

\textbf{Caption:}

\end{PromptBox}

\section{Captioning Metrics}
\label{sec:captioning-metrics}
We compute captioning metrics pairwise over aligned (reference, prediction) captions and average across the corpus.
\begin{itemize}
  \item \textbf{BLEU (corpus BLEU)}: n-gram precision with brevity penalty; 0--100 (higher is better). 
  \item \textbf{CHRF++ (CHRF)}: Character n-gram F-score (word order=2); 0--100 (higher is better).
  \item \textbf{ROUGE-L (F1)}: Longest common subsequence overlap (F1); 0--100 (higher is better).
  \item \textbf{BERTScore (F1)}: Semantic similarity via contextual embeddings; 0--100 (higher is better). \texttt{rescale\_with\_baseline=False}.
  \item \textbf{BGE-M3 Similarity}: Average cosine similarity of \texttt{BAAI/bge-m3} sentence embeddings; 0--100 (higher is better).
  \item \textbf{GPT-5 Rubric Similarity}: LLM-judged semantic agreement on a 0--5 rubric mapped to 0--100; higher is better.
\end{itemize}

\section{Data Licensing}

All SVG data used in this work originate from the SVG Stack~\citep{Rodriguez2025StarVector} and TheStack~\citep{kocetkov2022stack} datasets, therefore VectorGym data inherits the copyright and license characteristics of The Stack. SVG Stack is not an independent crawl of the web. It is a direct extraction of SVG files from The Stack~\citep{kocetkov2022stack}, the dataset maintained by the BigCode project. The Stack is a curated collection of source code repositories that have passed a strict license filtering pipeline. Only repositories under permissive licenses such as MIT, Apache, BSD, and CC0 are included, and repositories with non permissive or non redistributable licenses are excluded during collection.

The Stack also includes an opt out protocol that allows developers to request removal of their content. These removals are propagated automatically to all derived datasets. Since SVG Stack retains the original file paths and license identifiers from The Stack, it inherits the same governance and reflects all removals applied by BigCode.

Our work uses SVG Stack exactly as distributed, without adding external sources. All files therefore fall under permissive open source licenses that allow redistribution and research use. We intend to release the specific processed subset used in our experiments, which remains fully compatible with the original licensing terms.

\begin{table*}[t]
\centering
\caption{\textbf{VectorGym SVG Editing qualitative examples}. Results from models on the test set.}

\begin{tabular}{cc}
\includegraphics[width=0.48\linewidth]{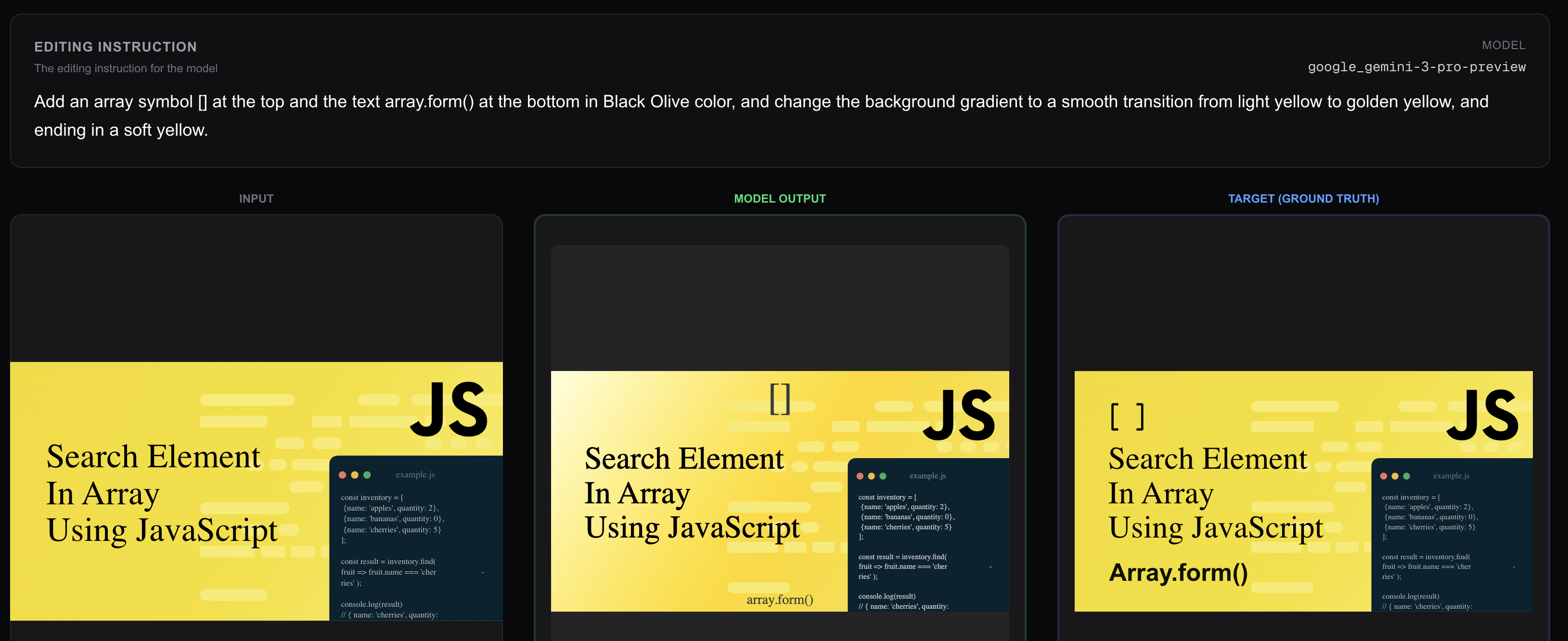} &
\includegraphics[width=0.48\linewidth]{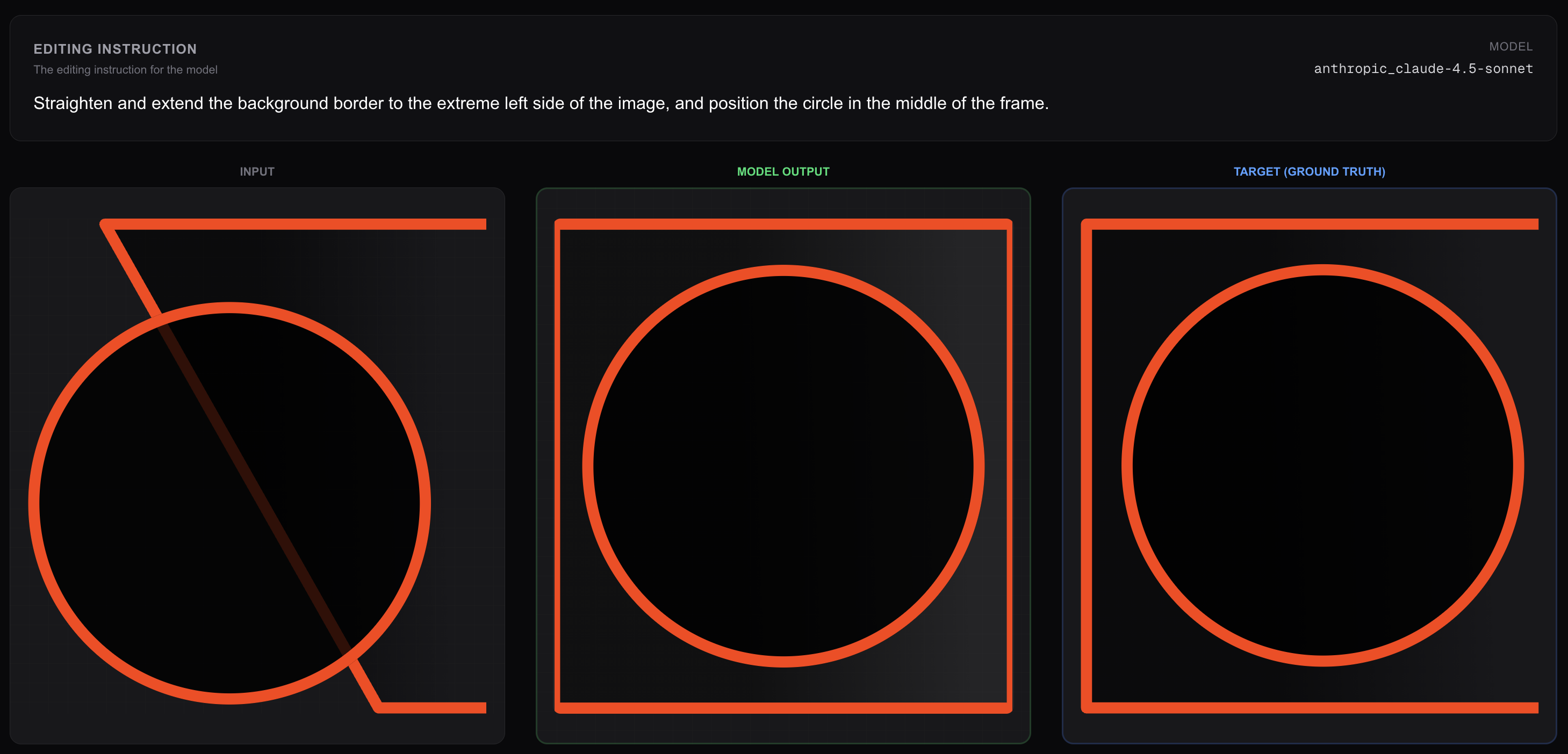} \\[6pt]
\includegraphics[width=0.48\linewidth]{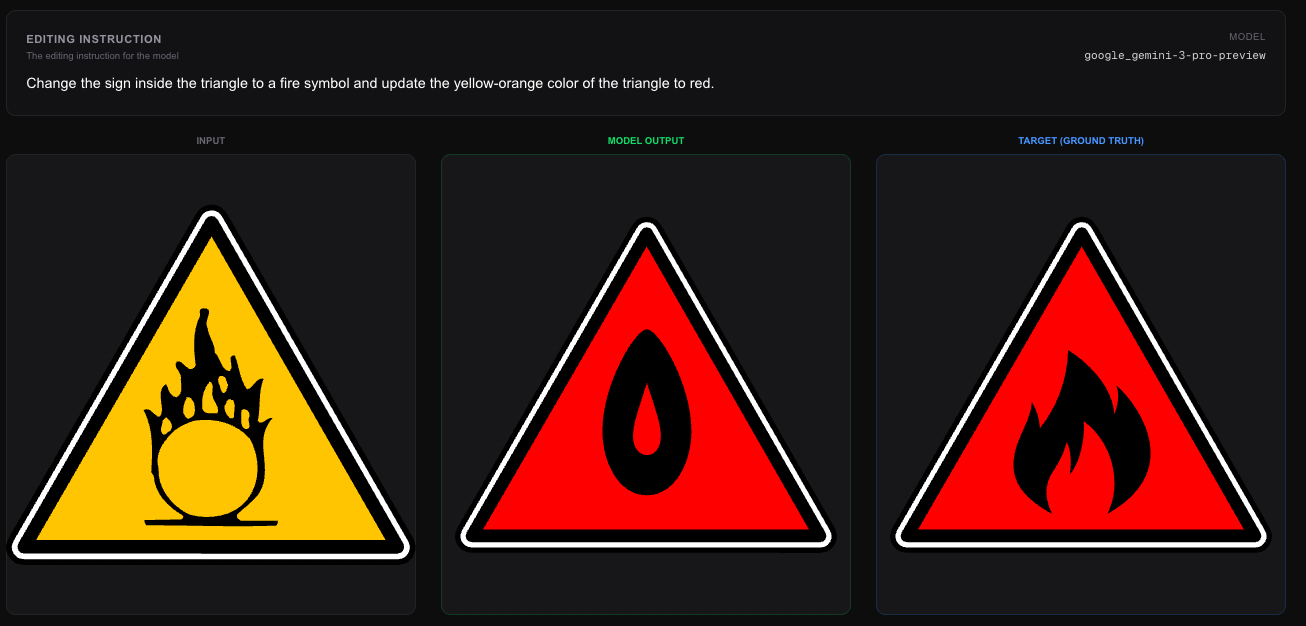} &
\includegraphics[width=0.48\linewidth]{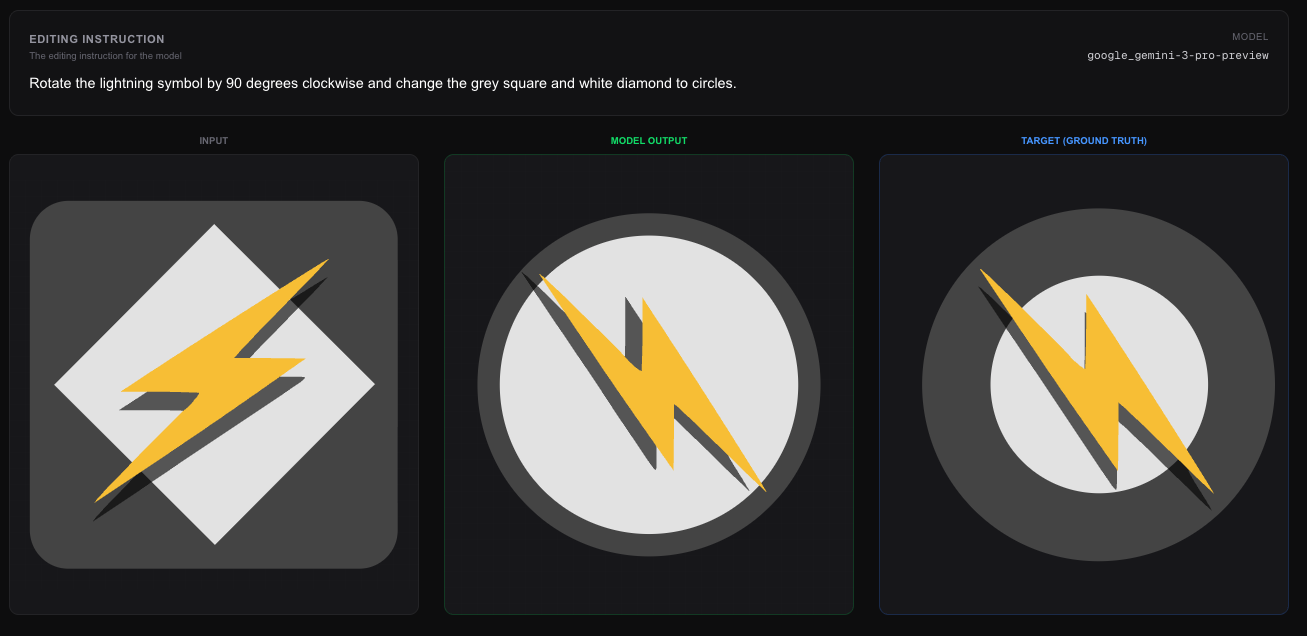}
\end{tabular}
\end{table*}

\begin{table*}[t]
\centering
\caption{\textbf{VectorGym Sketch-to-SVG qualitative examples}. Results from models on the test set.}

\begin{tabular}{cc}

\includegraphics[width=0.48\linewidth]{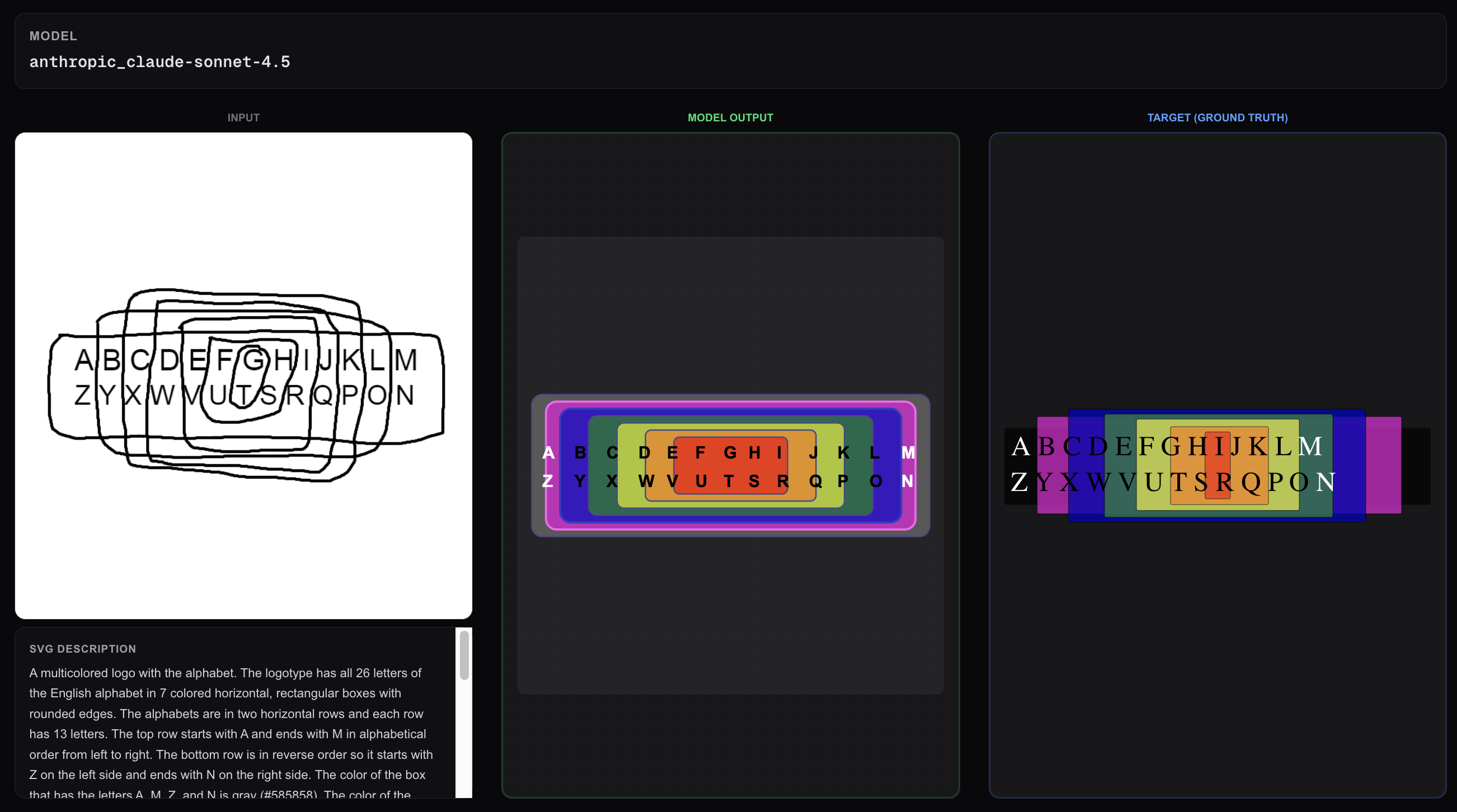} &
\includegraphics[width=0.48\linewidth]{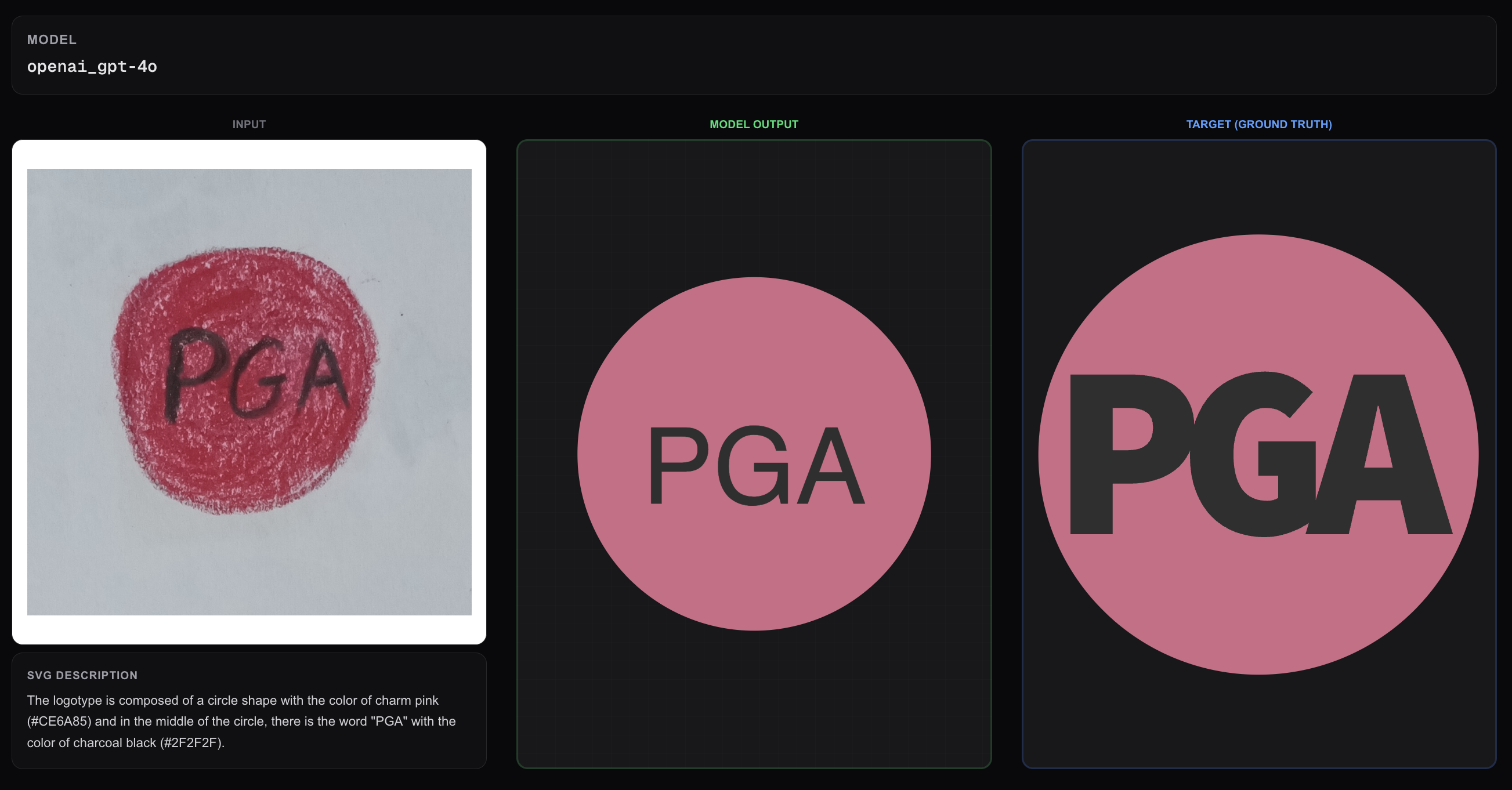} \\[6pt]
\includegraphics[width=0.48\linewidth]{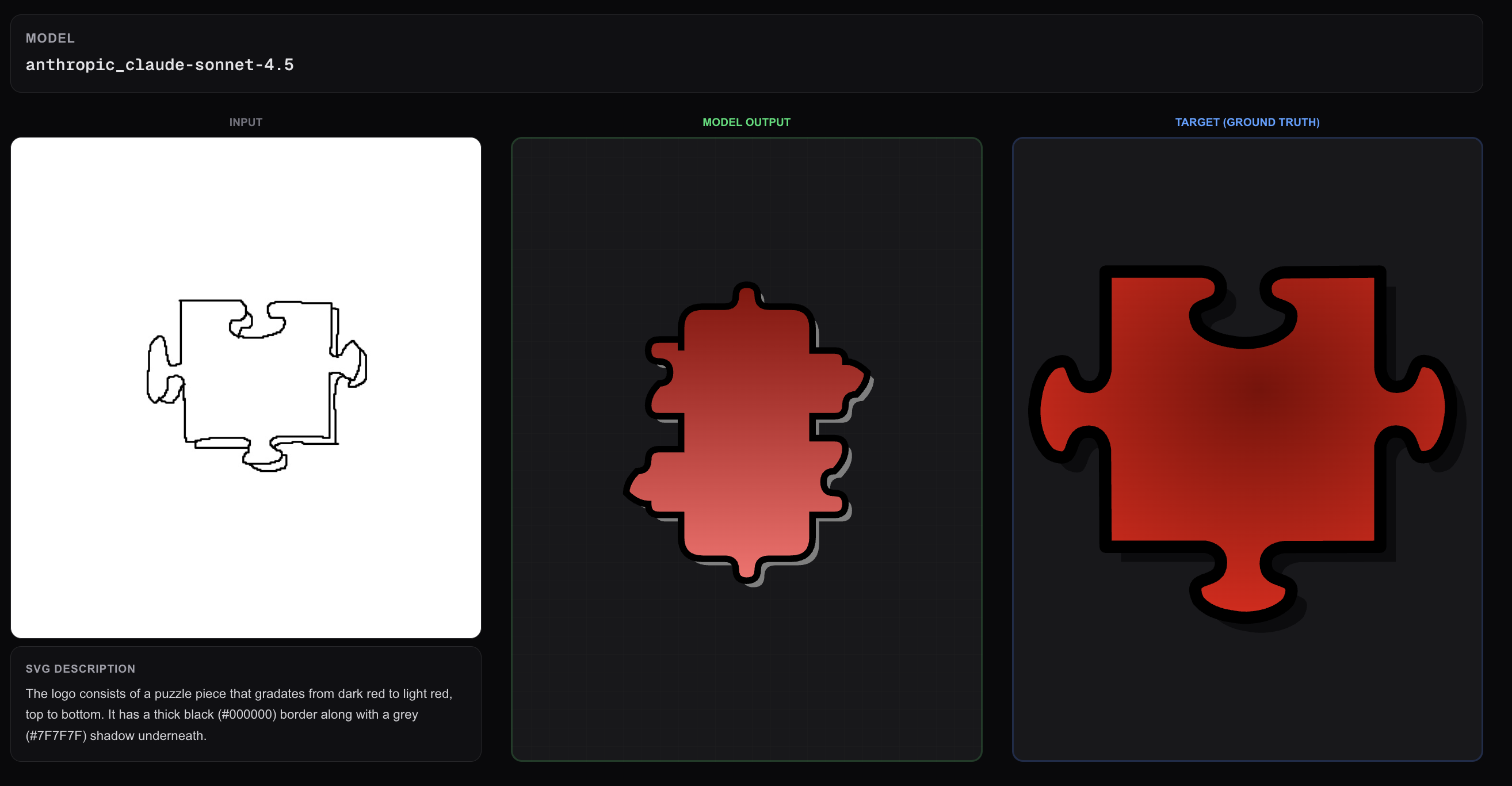} &
\includegraphics[width=0.48\linewidth]{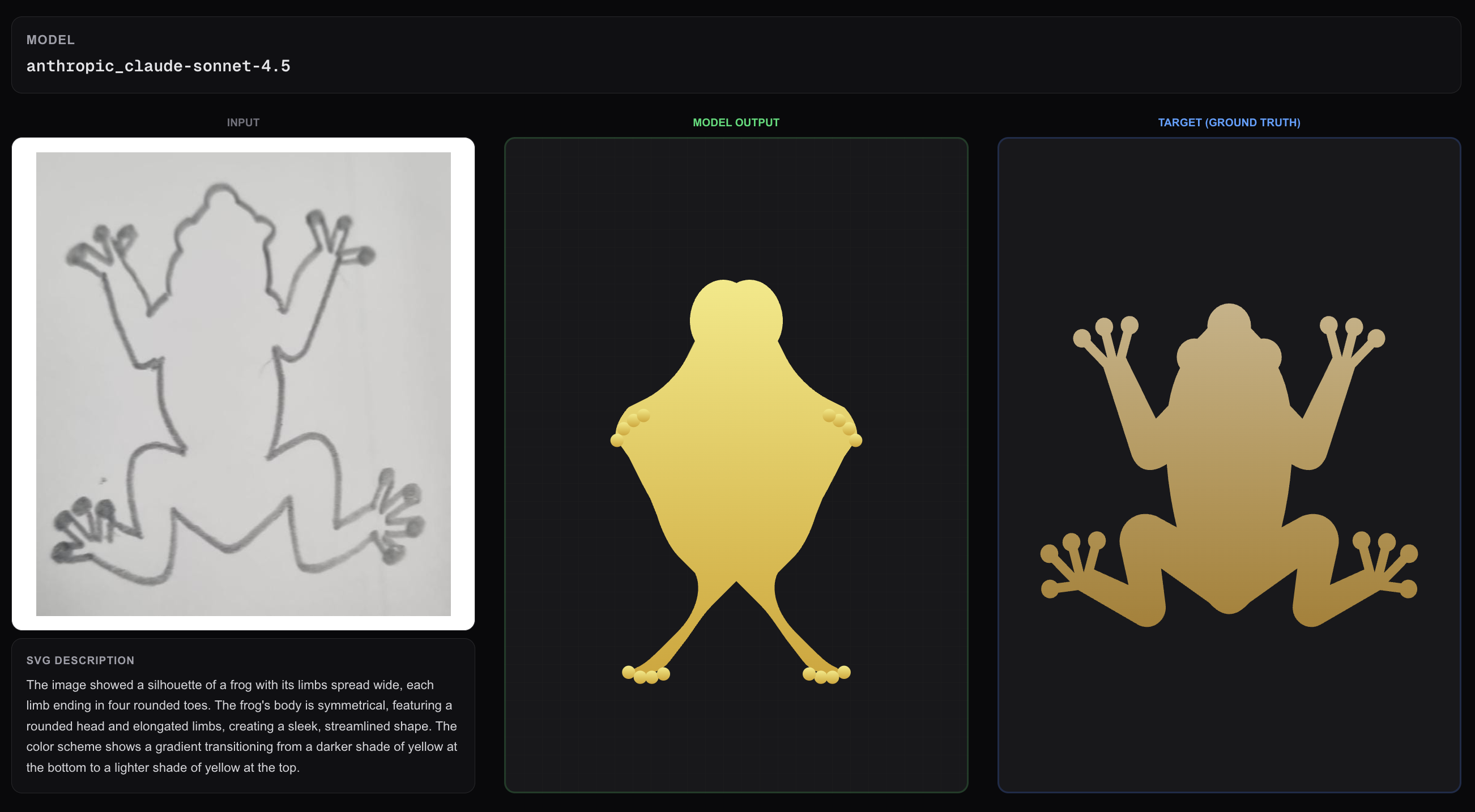}
\end{tabular}
\end{table*}

\begin{table*}[t]
\centering
\caption{\textbf{VectorGym Text-to-SVG qualitative examples}. Results from GPT4o on the test set.}

\begin{tabular}{cc}
\includegraphics[width=0.48\linewidth]{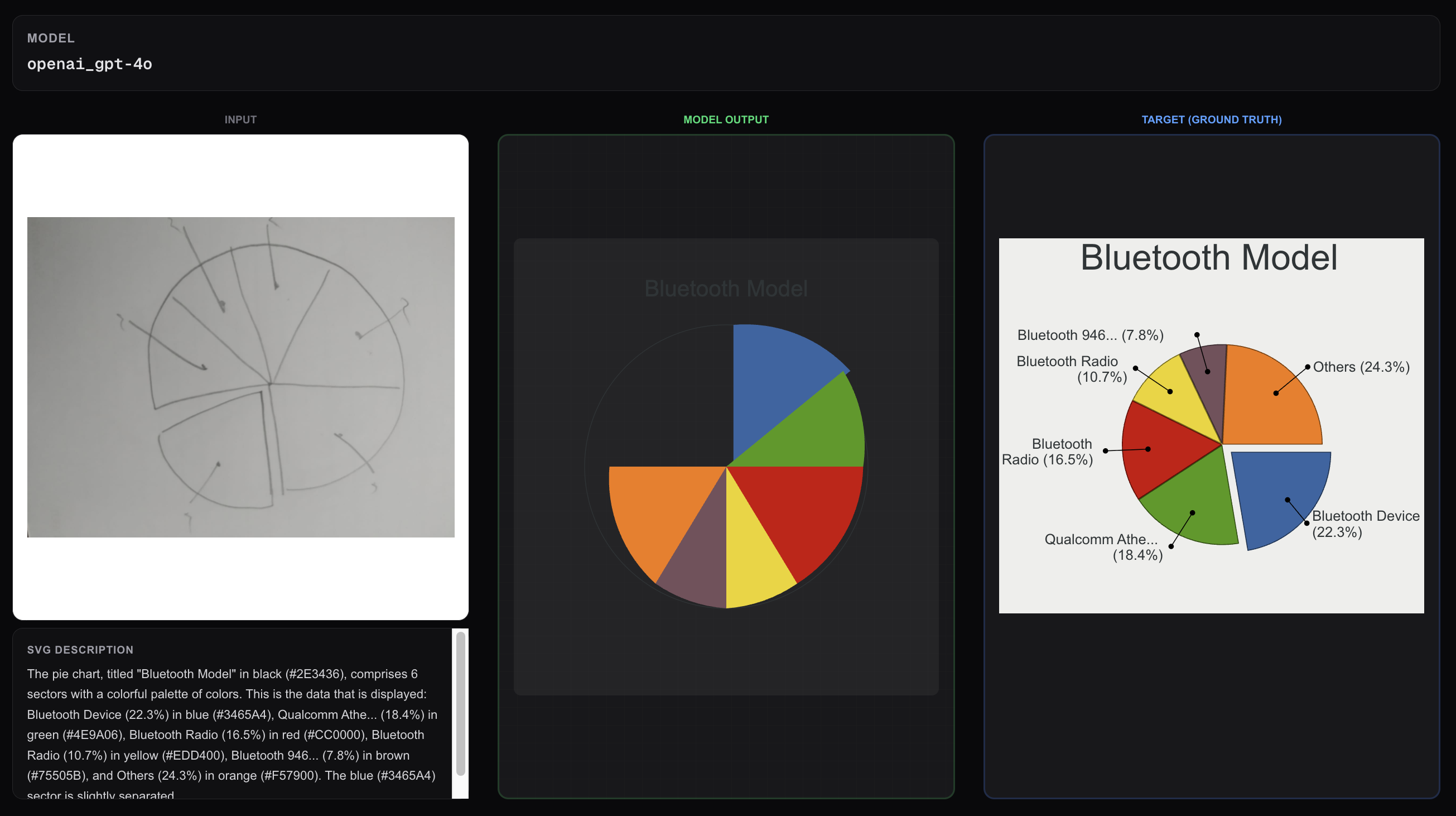} &
\includegraphics[width=0.48\linewidth]{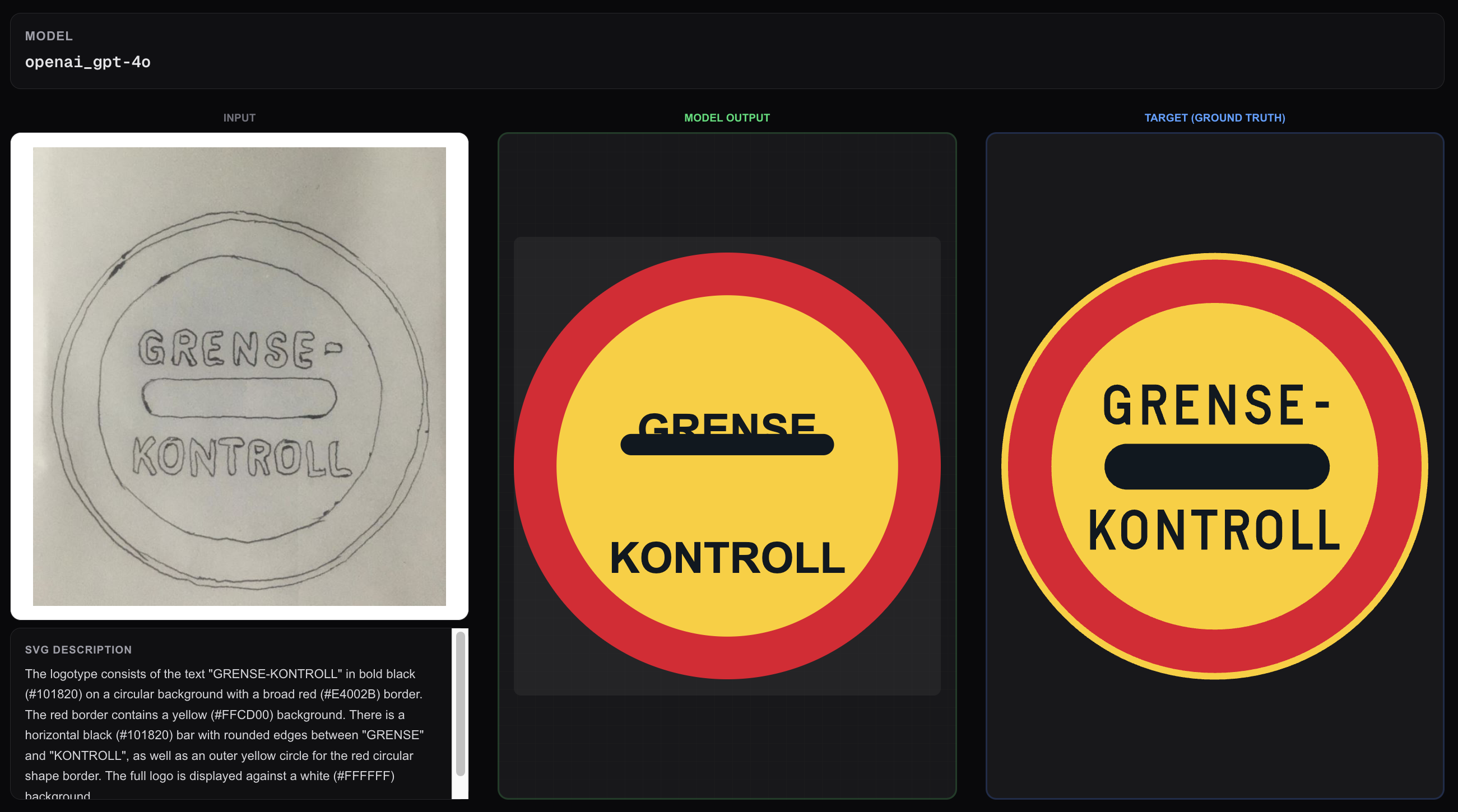} \\[6pt]
\includegraphics[width=0.48\linewidth]{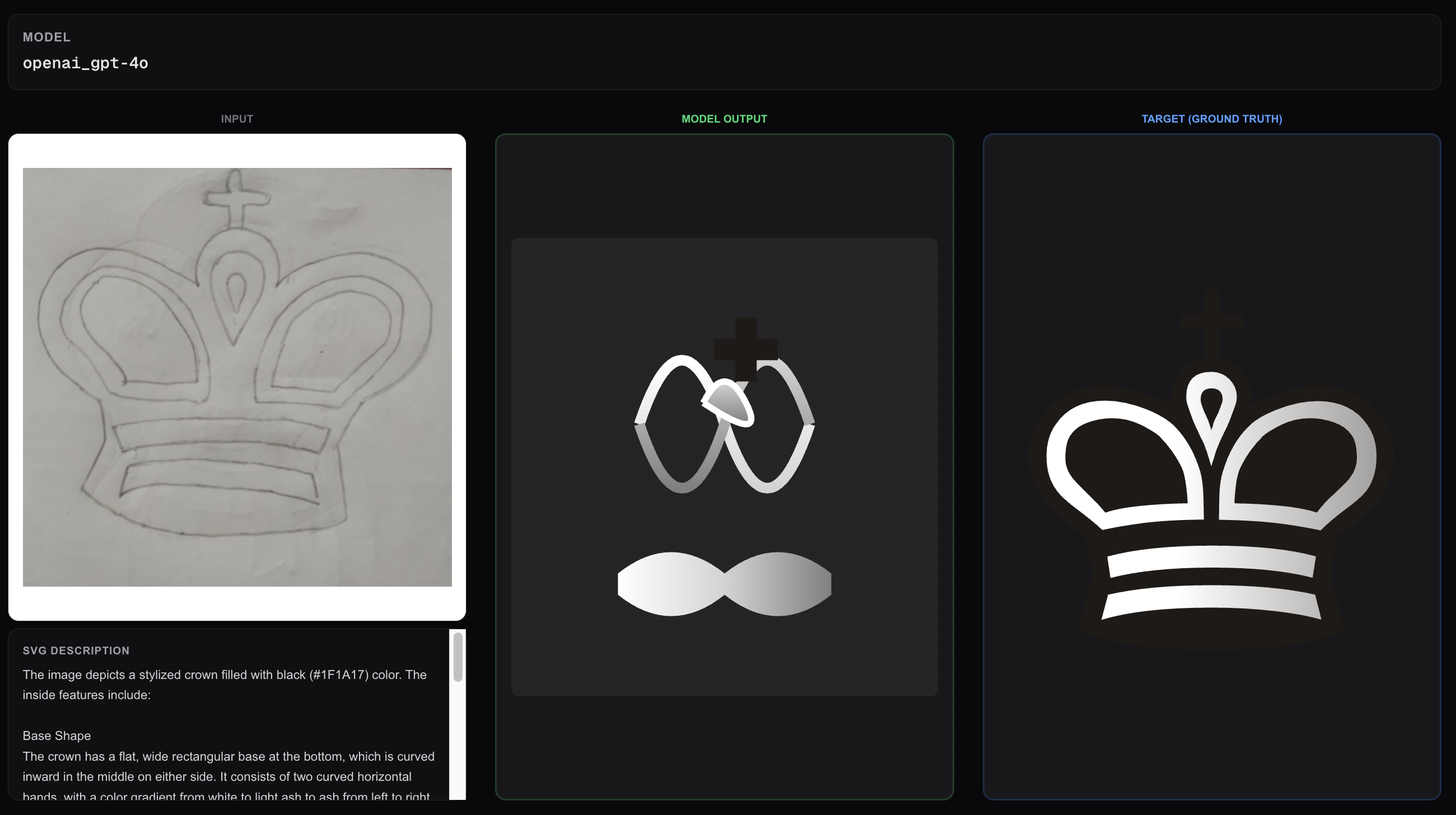} &
\includegraphics[width=0.48\linewidth]{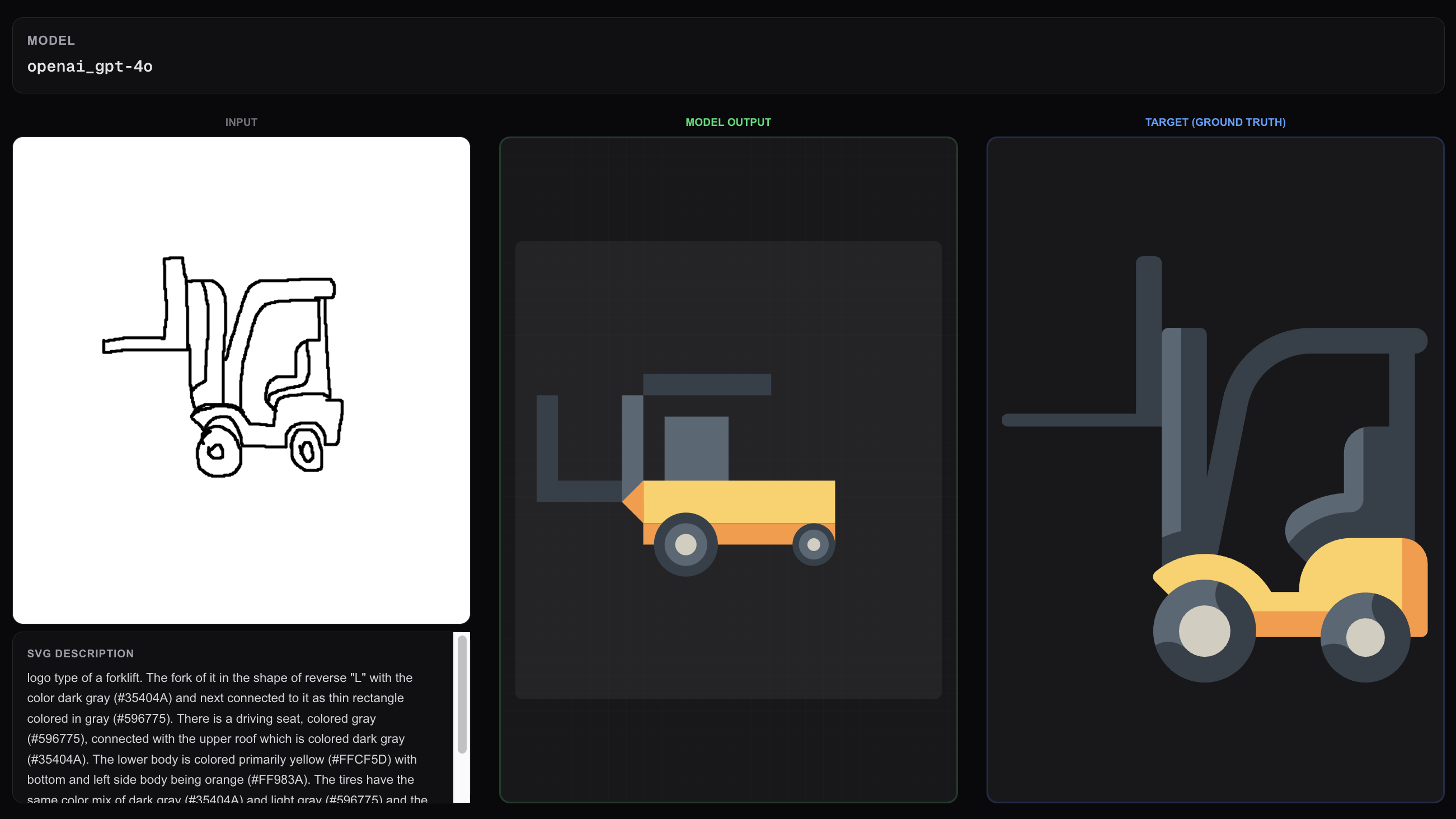}
\end{tabular}
\end{table*}

\begin{table*}[t]
\centering
\caption{\textbf{VectorGym SVG-Captioning qualitative examples}. Results from models on the test set.}

\begin{tabular}{cc}
\includegraphics[width=0.48\linewidth]{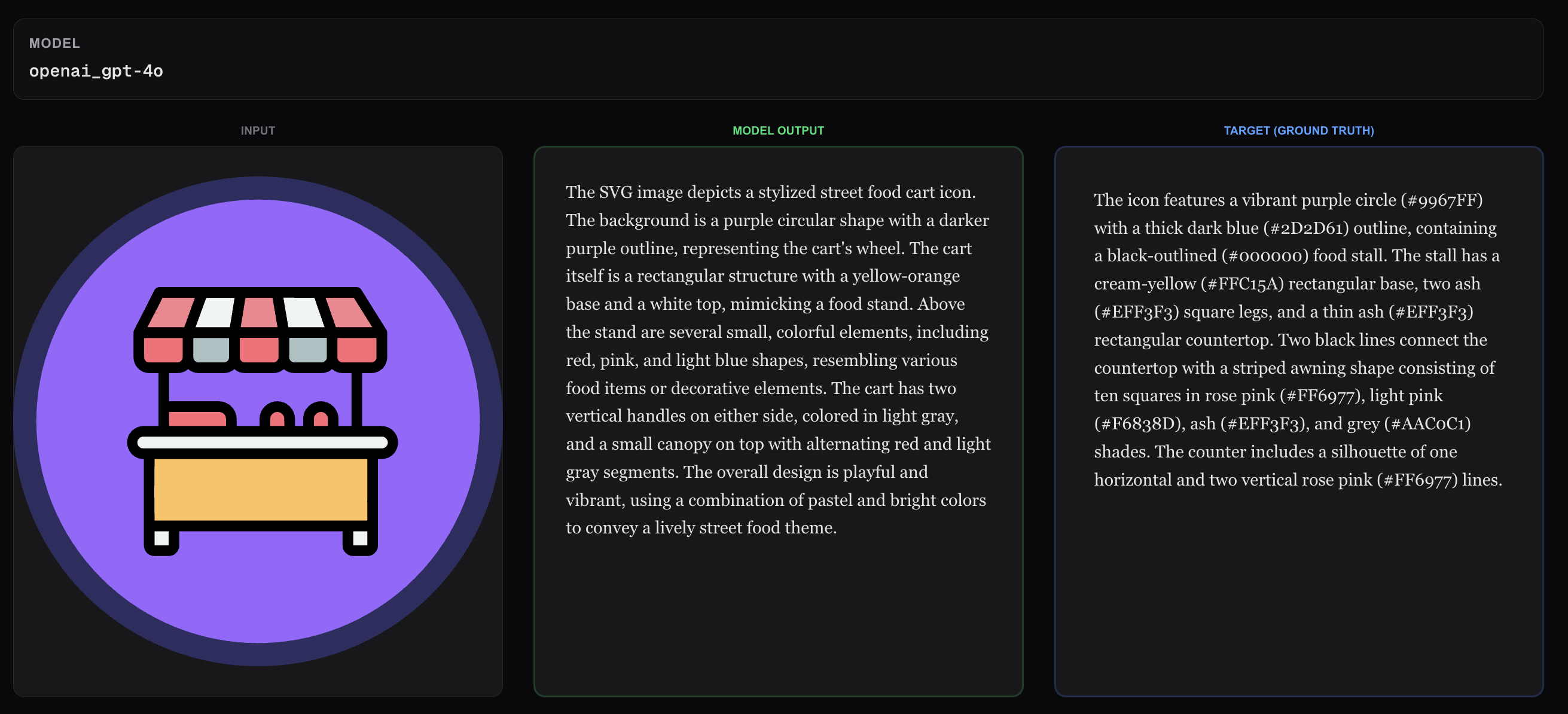} &
\includegraphics[width=0.48\linewidth]{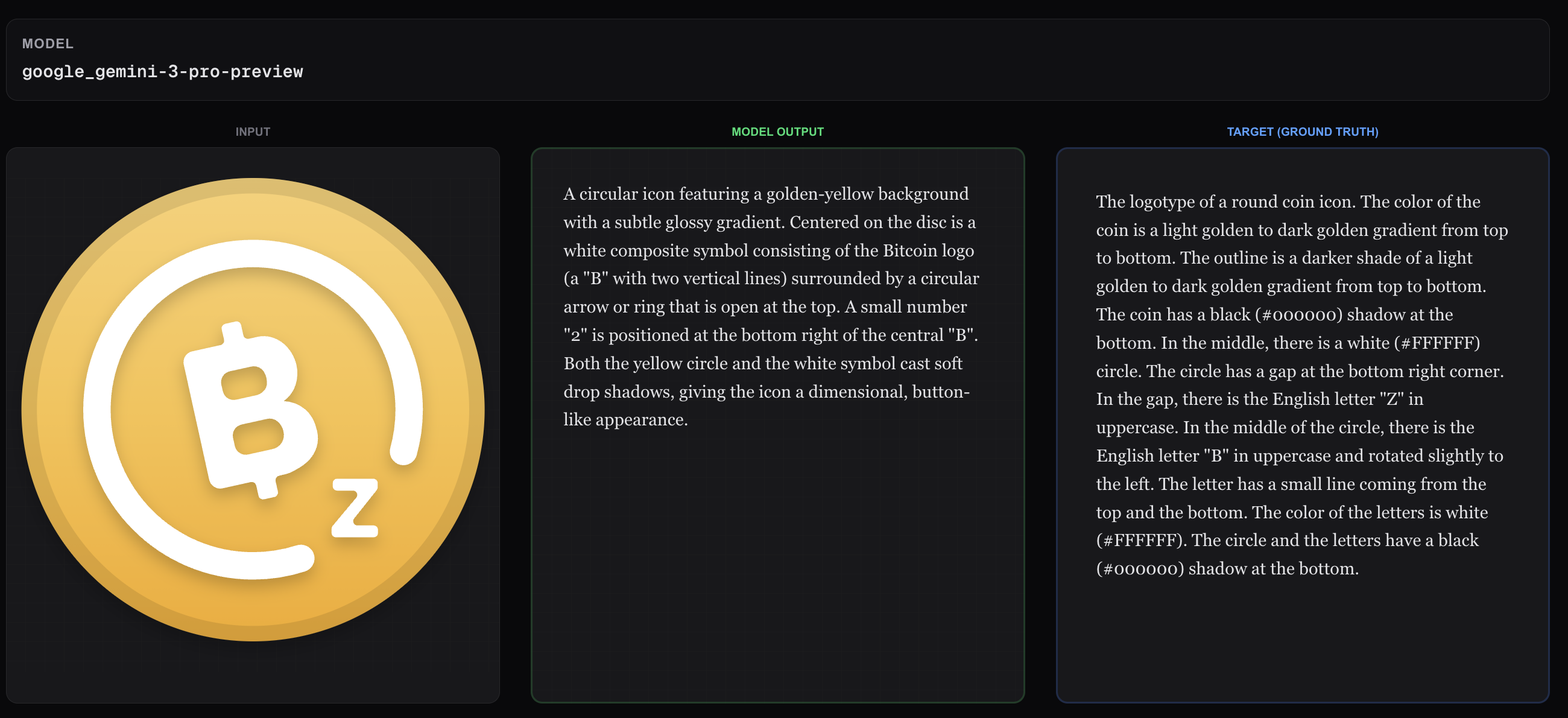} \\[6pt]
\includegraphics[width=0.48\linewidth]{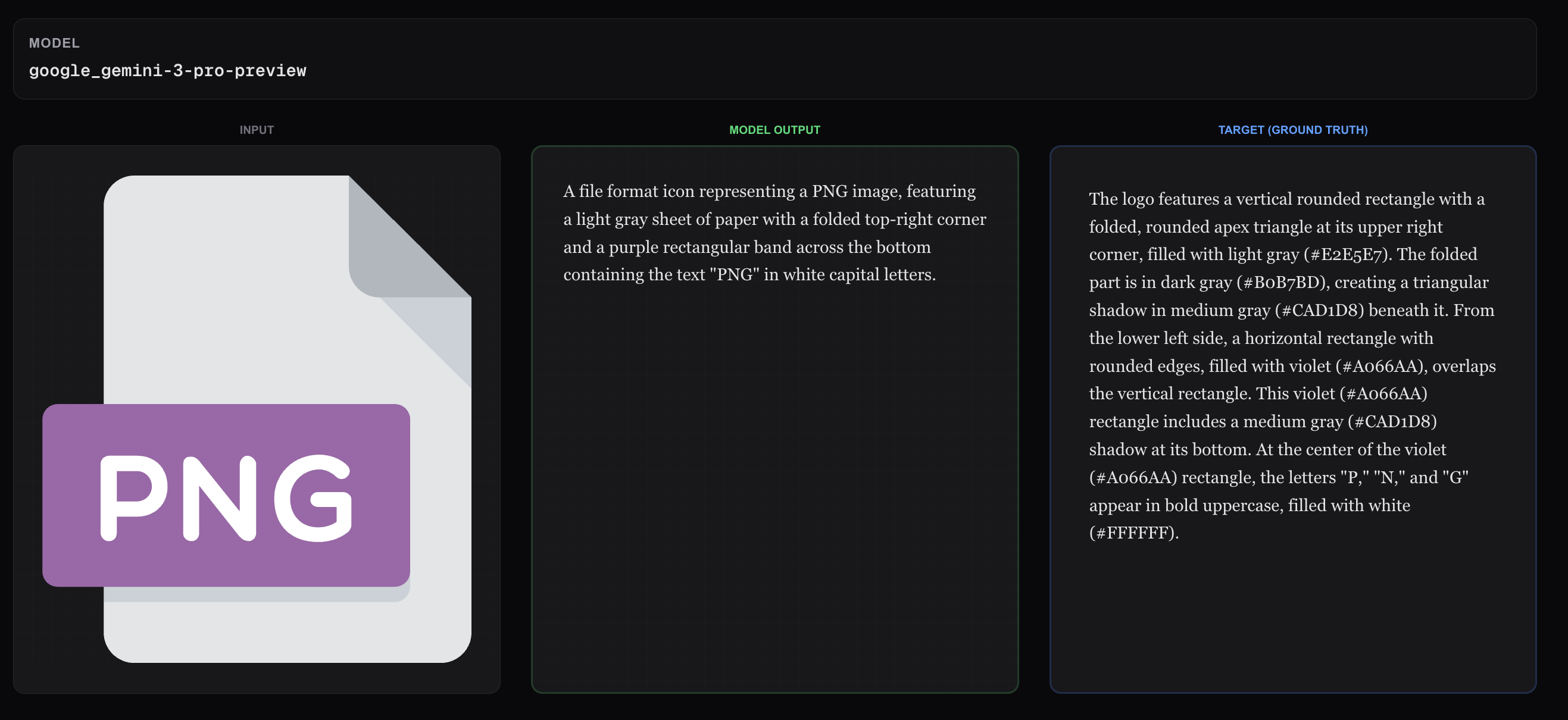} &
\includegraphics[width=0.48\linewidth]{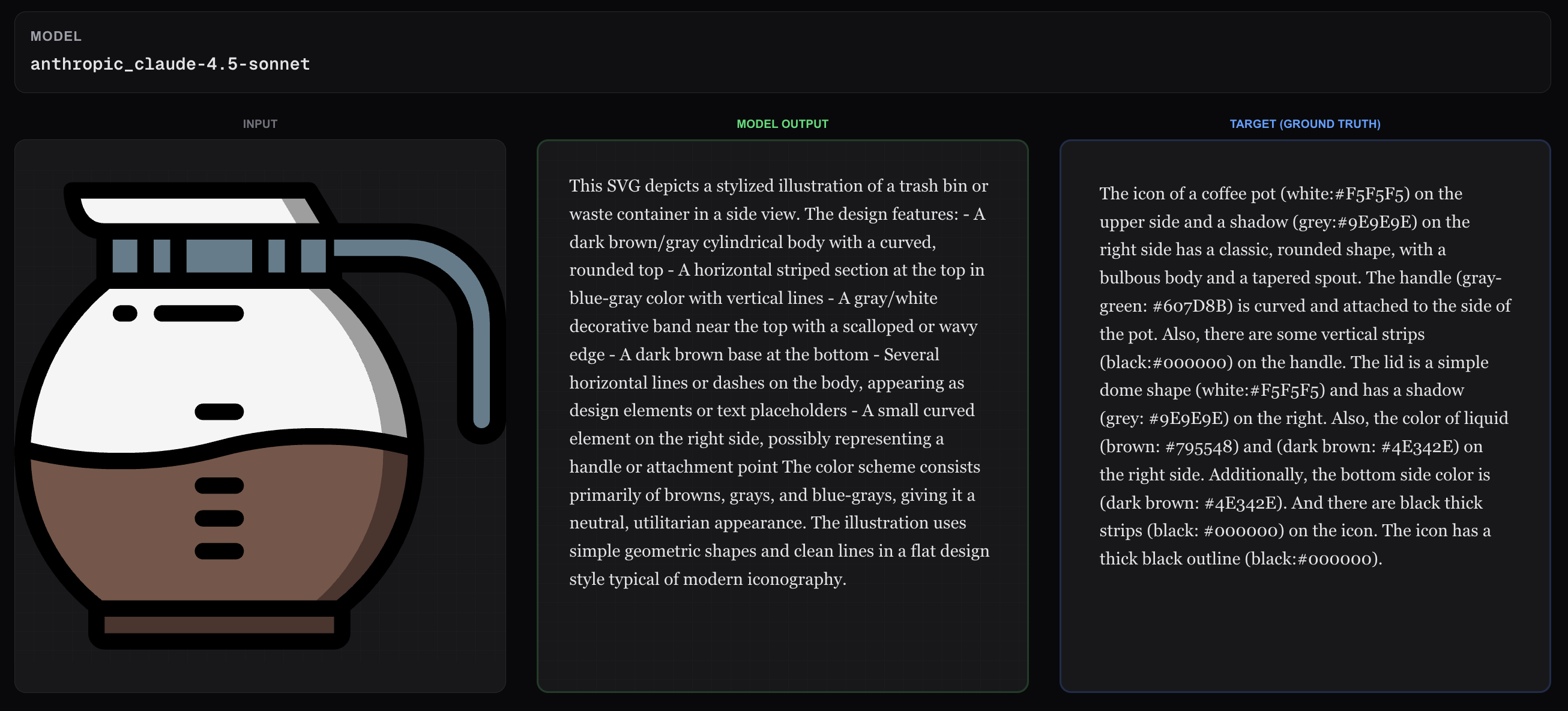}
\end{tabular}
\end{table*}
\end{document}